\documentclass{article}
\usepackage{arxiv}
\pdfoutput=1

\usepackage{microtype}
\usepackage{graphicx}
\usepackage{caption}
\usepackage{subcaption}
\usepackage{booktabs} 
\usepackage{amsmath}
\usepackage{multirow}
\usepackage{placeins}
\usepackage{hyperref}       
\usepackage{float}
\usepackage{breakcites}
\usepackage{natbib}
\usepackage{doi}
\usepackage[utf8]{inputenc} 
\usepackage[T1]{fontenc}    
\usepackage{hyperref}       
\usepackage{tcolorbox}
\usepackage{algorithm2e}
\usepackage{amsthm}

\usepackage{hyperref}
\usepackage{authblk}

\theoremstyle{definition}
\newtheorem{definition}{Definition}[section]

\title{Machine Learning Systems are Bloated and Vulnerable}

\author[1]{Huaifeng Zhang}
\author[1]{Fahmi Abdulqadir Ahmed}
\author[1]{Dyako Fatih}
\author[2]{Akayou Kitessa}
\author[3]{Mohannad Alhanahnah}
\author[1]{Philipp Leitner}
\author[4]{Ahmed Ali-Eldin}
\affil[1]{Chalmers University of Technology}
\affil[2]{Addis Ababa University}
\affil[3]{FitStack}
\affil[4]{Chalmers University of Technology and Elastisys AB}

\date{}

\renewcommand{\shorttitle}{Machine Learning Systems are Bloated and Vulnerable}

\begin{document}

\title{Machine Learning systems are Bloated and Vulnerable}

\maketitle

\begin{abstract}
    Today's software is bloated with both code and features that are not used by most users. This bloat is prevalent across the entire software stack, from operating systems and applications to containers.
Containers are lightweight virtualization technologies used to package code and dependencies, providing portable, reproducible and isolated environments.
For their ease of use, data scientists often utilize machine learning containers to simplify their workflow.
However, this convenience comes at a cost: containers are often bloated with unnecessary code and dependencies, resulting in very large sizes.
In this paper, we analyze and quantify bloat in machine learning containers.
We develop MMLB, a framework for analyzing bloat in software systems, focusing on machine learning containers. MMLB measures the amount of bloat at both the container and package levels, quantifying the sources of bloat. In addition, MMLB integrates with vulnerability analysis tools and performs package dependency analysis to evaluate the impact of bloat on container vulnerabilities. Through experimentation with 15 machine learning containers from TensorFlow, PyTorch, and Nvidia, we show that bloat accounts for up to 80\% of machine learning container sizes, increasing container provisioning times by up to 370\% and exacerbating vulnerabilities by up to 99\%.

\end{abstract}





\section{Introduction}\label{sec:intro}
Software technical debt is a metaphor introduced by Ward Cunningham in 1992 to describe long-term costs incurred during the software development process due to short-term workarounds that are meant to speed up the development process~\cite{cunningham1992wycash}. While technical debt can have solid technical reasons, this debt---like fiscal debt---needs to be serviced by, e.g., refactoring code, and removing unnecessary code (dead code) and unnecessary dependencies.
The technical debt of machine learning (ML) systems has recently come under
scrutiny~\cite{sculley2015hidden,tang2021empirical,obrien202223}. Sculley et al.~\cite{sculley2015hidden} identify eight main sources of technical debt in ML systems including common code issues such as dead code that is never used and duplicate code. The authors also observed that only a small fraction of code in ML systems in production is used for core machine learning functionality (training or serving).

\begin{figure}
    \centering
    \includegraphics[width=0.6\textwidth]{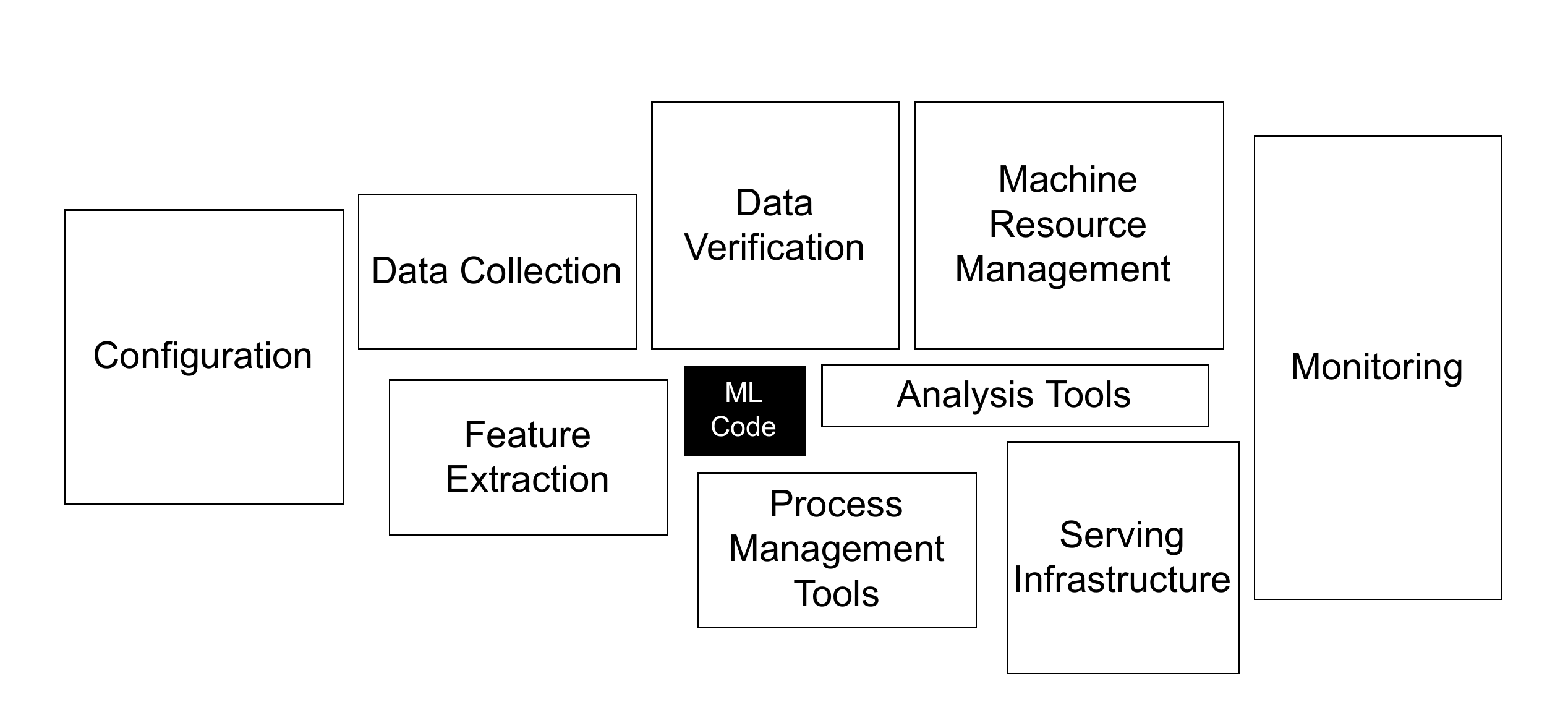}
    \caption{ML code is only a small part of ML production systems.}
    \label{fig:mlcode}
\end{figure}

Figure~\ref{fig:mlcode} shows functionalities included in real-world ML systems, where most of the code used in deployment is for ML-supporting functionalities.
To deploy an ML model, the most common industry practice is to use containers to package many of the required software components shown in the figure together~\cite{haller2022managing,googleCICD}.
Containers are designed to be lightweight, portable, and isolated environments that package an application and its dependencies together, providing a consistent and reproducible runtime environment.
However, in practice, containers can be bloated with unnecessary code and dependencies, resulting in very large sizes~\cite{zhao2020large,Rastogi2017Cimplifier,DockerSlim}.
Bloat can be defined simply as the parts of software that are never used during deployment. Software bloat is a result of many factors, including unnecessary software features~\cite{bhattacharya2013combining} and unnecessary code packaged with an application~\cite{soto2022coverage}.
Software bloat has been studied in the context of operating systems~\cite{kuo2020set}, containers~\cite{park2020toward}, Java applications~\cite{soto2021longitudinal}, among many other domains.

To the best of our knowledge, ML systems bloat has been mostly overlooked.
In this paper, we argue that the way ML containers are deployed, coupled with the technical debt in all the software components packaged in the container, leads to unnecessary bloat in ML system deployments that results in increased resource usage, decreased performance, and increased vulnerabilities.
We show that a significant portion of packaged code in ML systems results in unnecessary bloat that is now prevalent in ML deployments. ML containers are especially vulnerable to bloat as they aggregate the bloat in each component of the functionalities in Figure~\ref{fig:mlcode}, multiplying the effect of the bloat in the container.

Furthermore, the ML code, highlighted in black in Figure~\ref{fig:mlcode}, is especially prone to become bloated. Today most ML code runs using either TensorFlow~\cite{abadi2016tensorflow} or PyTorch~\cite{paszke2019pytorch}.  Many users rely on readily available models from, e.g., the TensorFlow Model Garden~\cite{TFModelGarden} coupled with, for example, the official TensorFlow container from DockerHub\footnote{https://hub.docker.com/}, or other similar model and container repositories~\cite{singh2021deploy,openja2022studying}. These containers come pre-packaged with all dependencies.
Many users use these containers as black boxes and do not further optimize the containers~\cite{zaharia2018accelerating,lee2018pretzel}. However, what people typically associate with ML code, i.e., the actual ML model, relies on only a small fraction of the code of these systems, with each ML model requiring different functionalities from the frameworks.
For example, the requirements from TensorFlow for model serving are different from those of model training.
Hence, many of the features relevant to model training, such as gradient calculation, will be bloat for a model serving deployment. ML deployments are always very specific, with different pipelines for training and serving~\cite{googleCICD,haller2022managing}. In this paper, we show how this setup results in unnecessary bloat in ML deployments, inheriting vulnerabilities in ML containers, and increasing the container resource usage and the bandwidth requirements for deployments while serving no useful purpose for the ML system.

To perform quantified measurements of bloat in ML containers, we develop a framework to debloat, analyze, and find vulnerabilities in ML containers. Our framework uses Cimplifier~\cite{Rastogi2017Cimplifier}, a tool for identifying bloat in ML containers. The framework then analyzes the bloat to find its source, its degree, the vulnerabilities the bloat introduces, and the package dependencies of bloat. To this end, we make the following five contributions.

\begin{itemize}

    \item We introduce MMLB, a framework for \emph{M}easuring and analyzing \emph{ML} deployment \emph{B}loat at the container level and the package level. In addition, MMLB quantifies vulnerabilities and performs package dependency analysis.
    \item We bridge the gap of existing research by being the first to conduct a comprehensive study of ML container bloat. While the framework can be used for other containers, the main focus of this work is ML containers.
    \item We study the bloat of 15 different containers running popular ML models using TensorFlow and PyTorch in training, tuning, and serving tasks. We quantify the storage overhead, the sources of bloat, and the increased provisioning latency caused by this bloat for these containers.
    \item We show that ML containers have many vulnerabilities. Bloat is responsible for 66\% to 99\% of all vulnerabilities in ML containers. While ML packages are the main source of bloat, generic packages are the main source of reported vulnerabilities in ML containers. 
    \item We open-source the framework\footnote{https://github.com/ChalmersMASS/MMLB}.
\end{itemize}

\section{Background}

In this section, we provide a brief background on the current state of software debloating, ML systems, and ML containers.

\subsection{Software Debloating}\label{sec:software_bloat}
Software bloat leads to increased resource consumption, decreased system security, and increased maintenance costs~\cite{qian2019razor}.
Bloat is a by-product of how software is developed today.
New features are regularly added to software for various reasons, such as addressing users' needs or attempting to attract new users.
However, not all of these new features will be successful, and thus not used by most users of the software.
Using containers simplifies the process of developing and deploying complex applications.
However, bloat is also added up vertically across the application in all layers when using containers.
New containers are typically built on top of existing containers, e.g., when creating a PyTorch container, a user will typically start from, for example, a Linux container on which they will install the PyTorch framework before publishing it to container registries like DockerHub. When new software is installed in a container, e.g., PyTorch, this is typically done using a software package manager such as Linux's Advanced Packaging Tool (APT)~\cite{APT} or Package Installer for Python (PIP)~\cite{PIP}. During installation, the package manager installs the software and all its dependencies.
Any bloat in these dependencies is also added to the ML container.

Software bloat has been studied extensively over the years~\cite{bhattacharya2011interplay, mitchell2009four,uhlig1995instruction} with recent focus from the security community~\cite{quach2019bloat,christensen2020decaf,azad2019less}. Reducing software bloat reduces the attack surface of an application, removing Common Vulnerabilities and Exposures (CVEs)~\cite{qian2019razor}, and gadgets~\cite{brown2019less}.
Due to the benefits of debloating, a large number of tools have been developed for removing software bloat. These tools can be divided into three main categories according to their granularity. The first set of tools focuses on debloating source code, e.g., LMCAS~\cite{LMCAS} and Chisel~\cite{heo2018effective}. The second set of tools focuses on debloating binaries, e.g., Nibbler~\cite{agadakos2019nibbler}. The third category of tools
focuses on file-level debloating, removing unused files from software. Examples of these tools include DockerSlim~\cite{gschwind2017optimizing} and Cimplifier~\cite{Rastogi2017Cimplifier}.

Source code and binary debloating tools remove unused code segments according to supplied program inputs. They are language-dependent, targeting applications written in a certain programming language, and are hence not suitable to use with most ML systems as ML software typically involves multiple programming languages. The file-level debloating tools are language-agnostic and can deal with different ML systems.
Therefore, in MMLB, we use a file-level debloating tool, Cimplifier with some modifications, to debloat ML containers.


\subsection{ML Systems}
In this paper, we aim to study how bloat affects ML systems.
While there are many frameworks for training and serving ML models, such as Caffe~\cite{jia2014caffe}, Keras~\cite{chollet2018keras}, MXNet~\cite{chen2015mxnet}, PyTorch~\cite{paszke2019pytorch}, and Tensorflow~\cite{abadi2016tensorflow}, the two most popular ones by downloads from DockerHub---the world's largest container registry---by far are TensorFlow and PyTorch. The official TensorFlow Docker container has been pulled more than 75.6 million times from DockerHub, while the official PyTorch container has been pulled more than 9.9 million times\footnote{Data obtained from DockerHub API(https://hub.docker.com/v2/repositories/<organization>/<image>) on January 4, 2024.}.
Due to their popularity, these systems have been actively developed with multiple releases and versions per year, a large developer community, and an active user base.

\begin{figure}[t]
    \centering
    \begin{subfigure}[b]{0.36\textwidth}
        \centering
        \includegraphics[width=\textwidth]{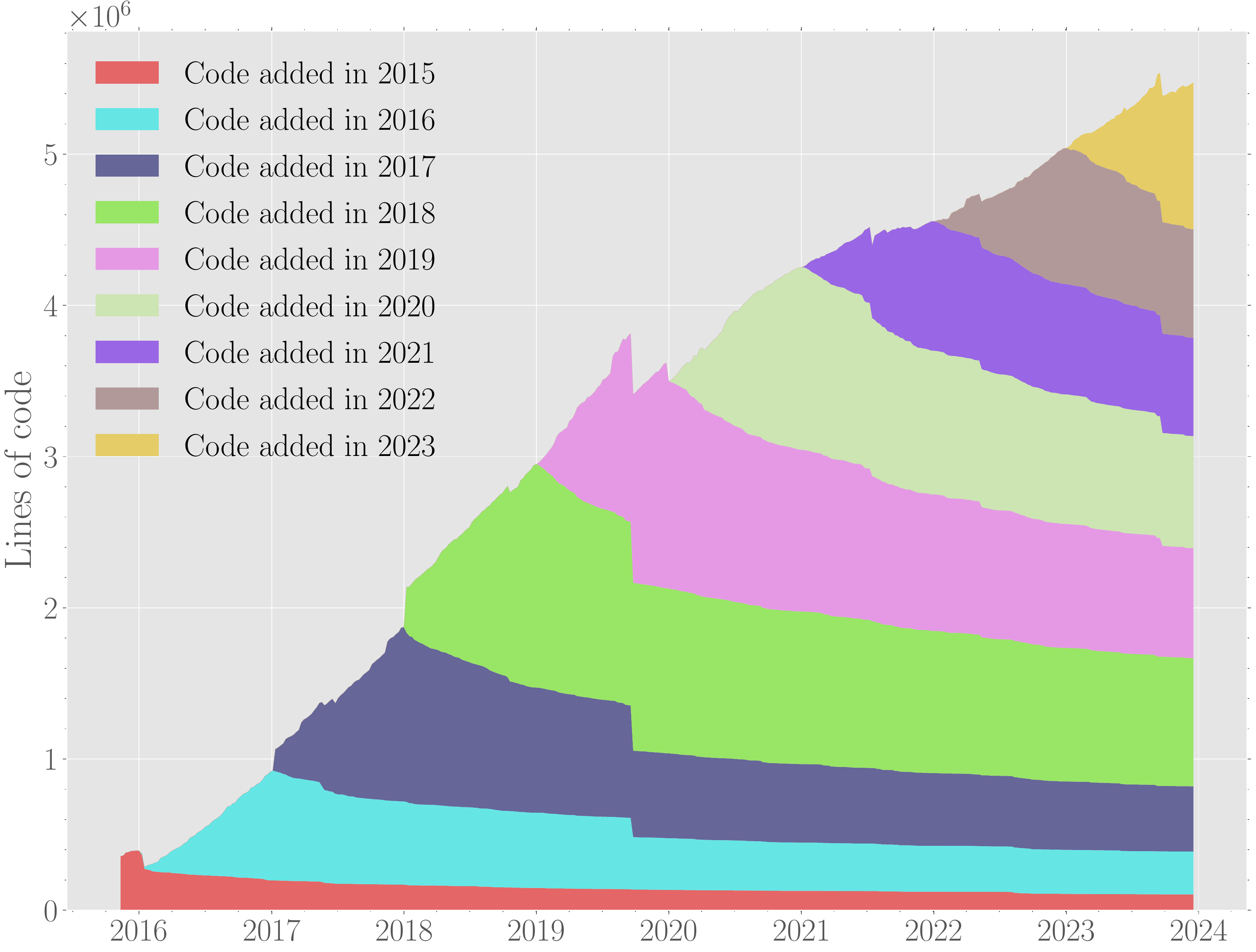}
        \caption{TensorFlow}
        \label{fig:TFLoC}
    \end{subfigure}
    \hfil
    \begin{subfigure}[b]{0.36\textwidth}
        \centering
        \includegraphics[width=\textwidth]{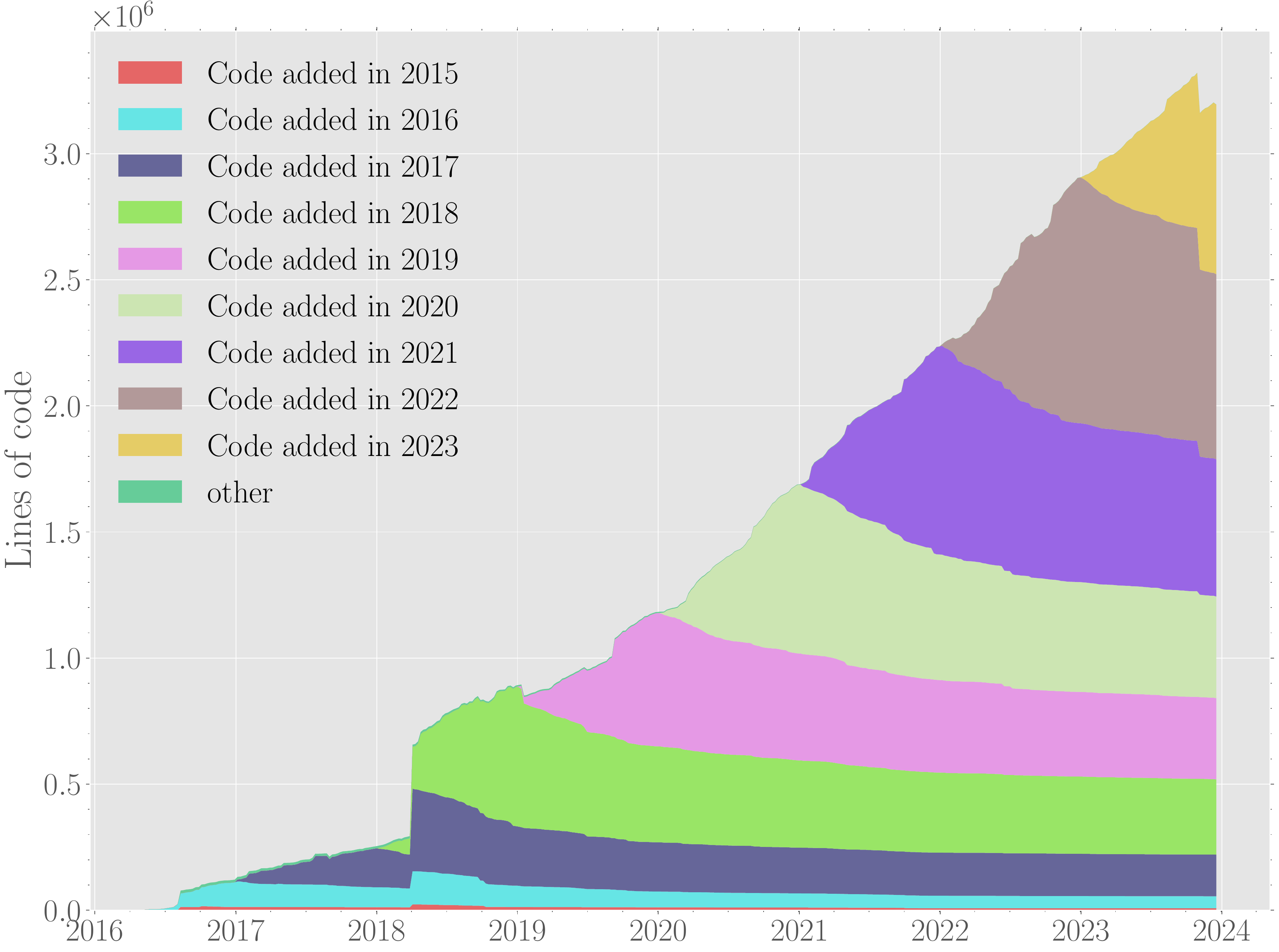}
        \caption{PyTorch}
        \label{fig:PTLoC}
    \end{subfigure}
    \caption{Stack-plots of the lines of code (LOC) added to TensorFlow and PyTorch over their lifetime. The $y-axis$ is the LoC. The $x-axis$ is time.
        The sudden drop in the LoC of TensorFlow in late 2019 is due to the release of TensorFlow 2.0, which removed a lot of code\protect\footnotemark.
        The large jump in the LoC of PyTorch in 2018 is due to the merging of PyTorch and Caffe2\protect\footnotemark.}
    \label{fig:LoC}
\end{figure}
\addtocounter{footnote}{-1}
\footnotetext[\thefootnote]{https://github.com/tensorflow/tensorflow/pull/31865}
\addtocounter{footnote}{1}
\footnotetext[\thefootnote]{https://github.com/pytorch/pytorch/issues/6032}

To visualize the code growth of TensorFlow and PyTorch over time, we use the Git-of-Theseus tool~\cite{Git} to analyze their Github repositories. Git-of-Theseus analyzes all Git commits to a GitHub repository to provide insights on a code base. We use the tool to create a stack plot of the code growth per year and how many Lines-of-Code (LoC) added in a year survive over time in the two projects as shown in Figure~\ref{fig:LoC}. The figure shows that both TensorFlow and PyTorch have been growing steadily over time, with TensorFlow and PyTorch adding on average 0.5 million and 0.4 million LoC per year, respectively. One can also see that while some code gets removed, most code that gets added to both projects remains unchanged.
Like most ML systems, PyTorch and TensorFlow are both written in C++ and Python. C++ is used to maximize computation performance, while the actual programming interfaces are written in Python for ease of use.

ML frameworks can run on CPUs.
However, due to the extensive computing workloads of ML tasks, ML frameworks mostly use hardware accelerators to accelerate ML computations. While there are many accelerators available today, the most commonly used accelerators are Nvidia GPUs. To use these GPUs, the ML frameworks use many libraries, mostly published by Nvidia, such as CUDA~\cite{CUDA}, CuDNN~\cite{CuDNN}, and CuBLAS~\cite{CuBLAS}. Many of these libraries were not originally designed for ML functionalities, but are now used as a core part of almost any ML program to run on GPUs.

\subsection{ML containers}\label{sec:ml_container}
Containers are lightweight virtualization technologies used to package code and dependencies in a virtual environment that can be easily migrated and redeployed across different clusters or computing environments. For their ease of use, containers are now the defacto deployment model used in ML production systems~\cite{googleCICD,haller2022managing}. Containers hide the complexity of ML systems by including all the different functionalities needed such as  CUDA~\cite{CUDA}, MKL~\cite{MKL}, analysis tools, feature extraction tools, and monitoring tools, in a single container.

Container technology uses a \emph{layered filesystem}\footnote{https://docs.docker.com/storage/storagedriver/}. In a layered filesystem, all files are organized in layers, with each layer representing a fully functioning container that is fully inherited by a new container that adds one more layer on top. Each layer has added files and folders that expand the functionality of the container, and in the process create a new container image. Hence, if any of these layers are bloated, the bloat will be inherited by any container that builds on top of the bloated layer.
The accumulation of bloat across layers contributes to the larger size of ML containers. To give an example, \textit{tensorflow/tensorflow:2.11.0-gpu}\footnote{https://t.ly/DYSuW}, a container image from TensorFlow hosted on DockerHub, has 39 layers with a total compressed size of 2.67GB. Any bloat in any of the 39 layers is inherited by the container.

\begin{figure}[htbp]
    \centering
    \begin{subfigure}[t]{0.49\textwidth}
        \centering
        \includegraphics[width=0.64\textwidth]{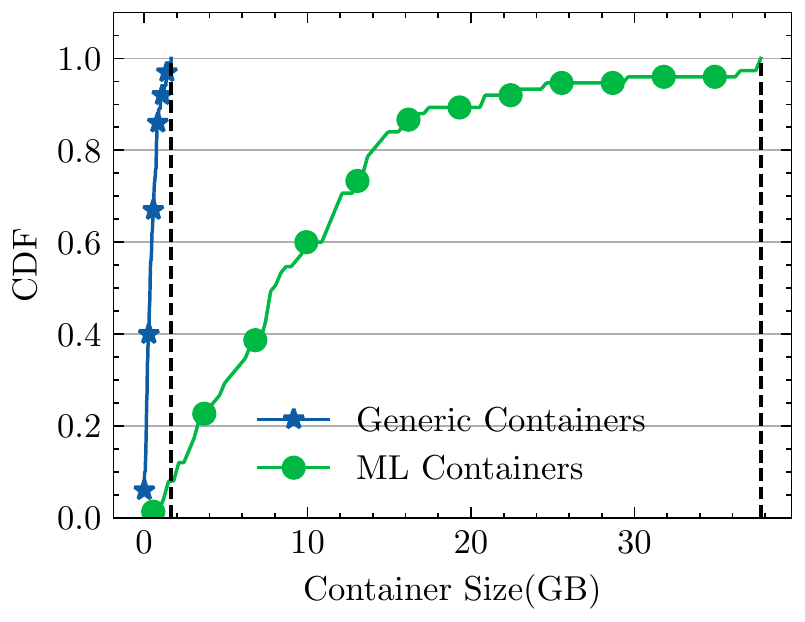}
        \caption{CDF of ML container sizes and generic container sizes.}
        \label{fig:img_size}
    \end{subfigure}
    \hfill
    \begin{subfigure}[t]{0.49\textwidth}
        \centering
        \includegraphics[width=0.675\textwidth]{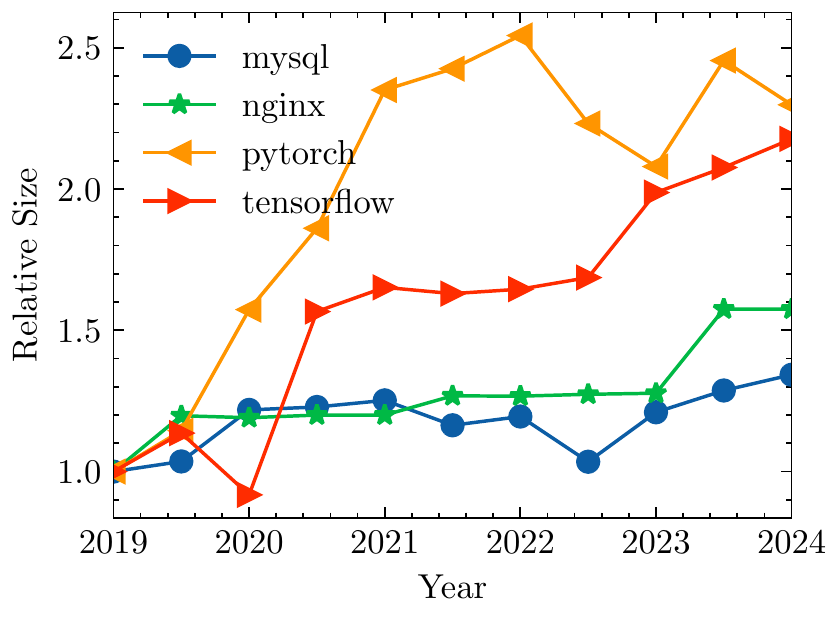}
        \caption{Increase in size of ML containers and generic containers.}
        \label{fig:size_evo}
    \end{subfigure}
    \caption{Sizes of ML containers versus generic containers. The cumulative distribution function (CDF) of container sizes is analyzed from the top 100 pulled generic containers and ML containers in DockerHub.}
    \label{fig:ml_vs_generic}
\end{figure}

Compared to other generic containers, the size of ML containers is on average larger than that of generic (i.e., non-ML) containers. To demonstrate this fact, we plot the sizes of the top 100 generic and ML containers pulled from DockHub, and the growth of four example containers over the years in Figure~\ref{fig:ml_vs_generic}.
The size of ML containers is at least an order of magnitude larger than that of generic containers as can be seen in Figure~\ref{fig:img_size}.
In Figure~\ref{fig:size_evo}, 
the increase in the sizes of basic ML containers, with only the TensorFlow and PyTorch frameworks, is substantial, doubling over the past five years.
Although the size of the PyTorch container dropped during 2022 because a package named \texttt{cudatoolkit} was removed from the container, the size of the container increased again during 2023.
However, generic containers like \texttt{nginx} and \texttt{mysql}, two of the most pulled generic containers in DockerHub, have only experienced minor growth.
The large size of ML containers leads to longer provisioning times that can be prohibitive for some applications, driving research on how to optimize ML container provisioning for, e.g., edge applications~\cite{chen2022starlight,fu2020fast, gujarati2017swayam}. Hence, it is of paramount importance to understand how ML containers are bloated, as reducing the size of these containers can result in substantial bandwidth and storage savings for hosting and deployments of these containers while decreasing their attack surface and vulnerabilities.


\section{Experiment setup and Framework}
\begin{figure}
    \centering
    \includegraphics[scale=0.36]{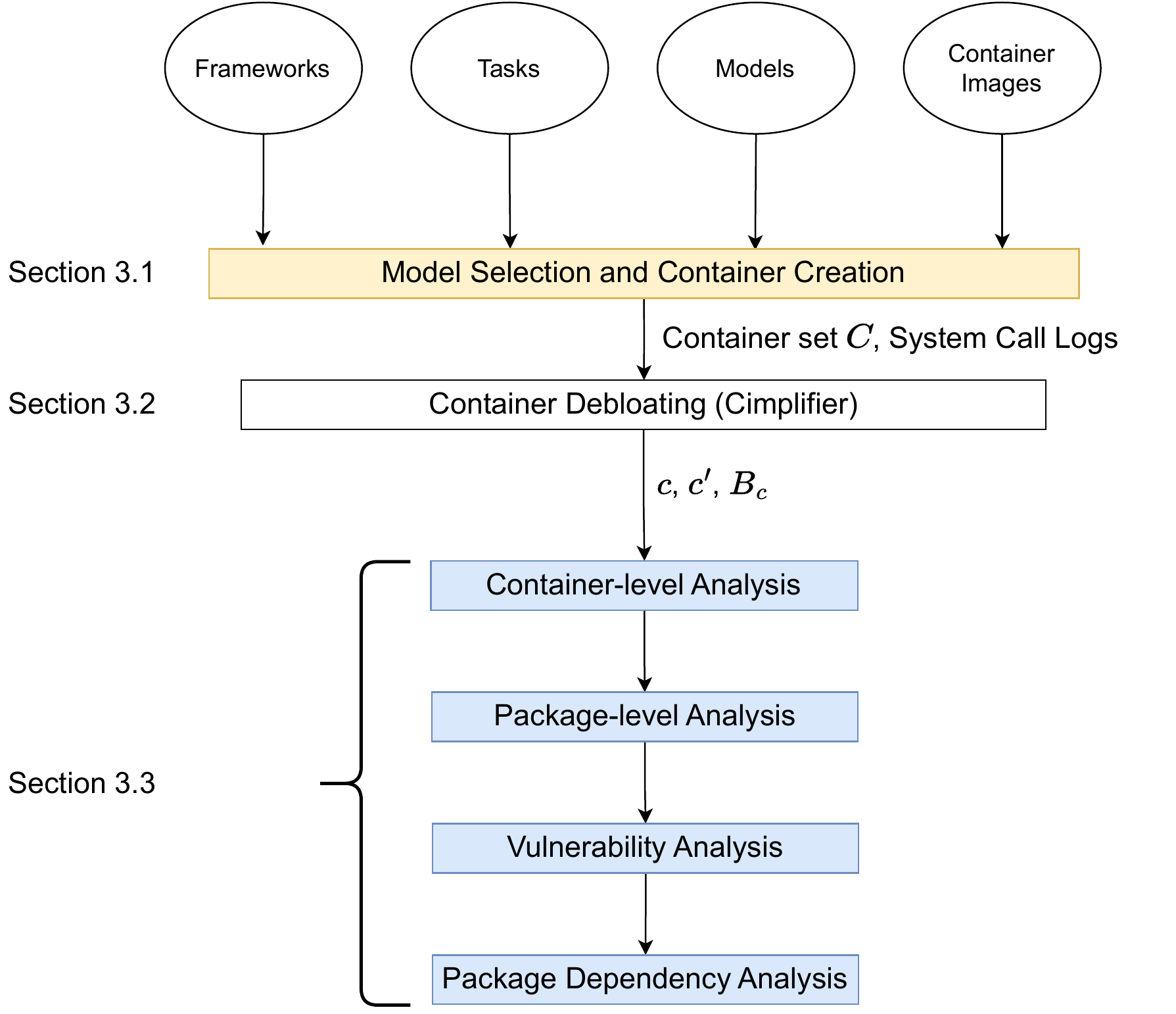}
    \caption{MMLB Framework overview. MMLB selects various combinations of ML frameworks, ML tasks, ML models, and ML container images. MMLB then debloats the containers using a modified version of Cimplifier. The debloated containers are then analyzed using container-level analysis, package-level analysis, vulnerability analysis, and package dependency analysis to find the bloat degree, the sources of bloat, the vulnerabilities, and the package dependencies.}
    \label{fig:framework}
\end{figure}

We now introduce our framework for removing and analyzing bloat in ML containers. The main components of our framework are shown in Figure \ref{fig:framework}. As our goal is to evaluate bloat in ML systems, we start the section with a detailed description of how we select the evaluated containers. We delve into the details of each stage in the framework next.

\subsection{Model Selection and Container Creation}
\label{sec:container_creation}


As there is an ever-increasing number of ML models and ML frameworks, for our analysis to be representative, we select various combinations of ML frameworks, ML tasks, ML models, and ML container images to examine. Table~\ref{tab:containers} summarizes the combinations we used in our experiments.
We focus our study on TensorFlow and PyTorch, as they are the most widely used ML frameworks considering the number of downloads from DockerHub. 

There are three main tasks for ML systems, namely training, tuning and serving. In addition, ML models are used across various domains, including Natural Language Processing (NLP), Image Classification (IC), and Image Segmentation (IS). For each domain, there are many developed models, and each of these models activates different parts of an ML system. To better understand this, our experiments evaluate the bloat of multiple popular models for all three tasks and for different ML domains. We use TensorFlow Serving~\cite{TFServing} and TorchServe~\cite{TorchServe} for serving. We sourced models from open-source model repositories, including TensorFlow Model Garden~\cite{TFModelGarden}, PyTorch Examples~\cite{PTExamples}, Fairseq~\cite{Fairseq}, and Nvidia DLExamples~\cite{Dlexpamples}. We also selected container images from the official Docker TensorFlow (TF) and PyTorch (PT) images, as well as Nvidia GPU Cloud (NGC) images. In total, we evaluate 15 containers (see Table \ref{tab:containers}).

\begin{footnotesize}
    \begin{table*}[htb]
        \begin{center}
            \caption{The containers studied in this paper.}
            \label{tab:containers}
            \scalebox{0.9}{
            \begin{tabular}{llllllll}
                \toprule
                Container & Framework        & Task     & Model        & Image                           & Publisher & Application & Year \\
                \midrule
                $c_1$     & TensorFlow-2.4.0 & tuning   & Bert         & tensorflow/tensorflow:2.4.0-gpu & TF        & NLP         & 2020 \\
                $c_2$     & PyTorch-1.12.0   & tuning   & Bert         & pytorch/pytorch:latest          & PT        & NLP         & 2022 \\
                $c_3$     & PyTorch-1.11.0   & tuning   & Bert         & bert:latest                     & NGC       & NLP         & 2022 \\
                $c_4$     & TensorFlow-2.4.0 & tuning   & Bert         & bert\_tf2:latest                & NGC       & NLP         & 2020 \\
                $c_5$     & PyTorch-1.11.0   & serving  & Bert         & pytorch/torchserve:latest-gpu   & PT        & NLP         & 2022 \\
                $c_6$     & TensorFlow-2.4.0 & serving  & Bert         & tensorflow/serving:2.4.0-gpu    & TF        & NLP         & 2020 \\
                $c_7$     & TensorFlow-2.4.0 & training & Transformer  & tensorflow/tensorflow:2.4.0-gpu & TF        & NLP         & 2020 \\
                $c_8$     & PyTorch-1.12.0   & training & Transformer  & pytorch/pytorch:latest          & PT        & NLP         & 2022 \\
                $c_9$     & TensorFlow-2.6.0 & training & EfficientNet & efficientnet\_v2\_tf2:latest    & NGC       & IC          & 2021 \\
                $c_{10}$  & TensorFlow-2.4.0 & training & ResNet       & tensorflow/tensorflow:2.4.0-gpu & TF        & IC          & 2020 \\
                $c_{11}$  & PyTorch-1.12.0   & training & ResNet       & pytorch/pytorch:latest          & PT        & IC          & 2022 \\
                $c_{12}$  & PyTorch-1.11.0   & serving  & ResNet       & pytorch/torchserve:latest-gpu   & PT        & IC          & 2022 \\
                $c_{13}$  & TensorFlow-2.4.0 & serving  & ResNet       & tensorflow/serving:2.4.0-gpu    & TF        & IC          & 2020 \\
                $c_{14}$  & TensorFlow-2.4.0 & training & MaskRCNN     & tensorflow/tensorflow:2.4.0-gpu & TF        & IS          & 2020 \\
                $c_{15}$  & TensorFlow-2.4.0 & training & MaskRCNN     & nvidia\_mrcnn\_tf2:latest       & NGC       & IS          & 2020 \\
                \midrule
                \multicolumn{8}{c}{NLP = Natural Language Processing; IC = Image Classification; IS = Image Segmentation.} \\
                \bottomrule
            \end{tabular}
            }
        \end{center}
    \end{table*}    
\end{footnotesize}

\subsection{Container Debloating} \label{sec:debloat}


ML systems are built mostly using C++ and Python, and delivered as shared libraries and Python packages. Container debloating tools such as DockerSlim~\cite{DockerSlim} and Cimplifier~\cite{Rastogi2017Cimplifier} are the only suitable set of tools for debloating such systems that were implemented with multiple programming languages. 
We ran experiments with both DockerSlim and Cimplifier. In our experiments, DockerSlim removes many needed dynamically linked ML GPU libraries, resulting in broken containers.

The second tool, Cimplifier~\footnote{We obtained the original code for Cimplifier from the authors. Both our team and the Cimplifier team collaborated to open-source it as part of MMLB.}, has more than just debloating functionalities and can be used to partition a container into multiple containers. We only evaluated the debloating feature of Cimplifier. To debloat, Cimplifier takes system call logs extracted from the container based on a given workload as input and identifies the files accessed using the system calls. Accessed files are then considered necessary for the container's functioning for the given workload, while other files are considered bloat and removed. 
In this way, Cimplifier can remove both package files and non-package files.
The output is a debloated container with only files necessary for the workload.

Similar to DockerSlim, Cimplifier also failed to debloat some containers due to a problem in the system call log extraction workflow which was unable to identify some of the files accessed with more complex workloads. In our development of MMLB, we modified Cimplifier to parse more complex system call logs, which enables Cimplifier to correctly identify accessed files with complex workloads.
We also removed irrelevant features and only retained the debloating-related features. 
Additionally, we automated the process of collecting system call logs to streamline the workflow of Cimplifier and MMLB.

The system call logs and the container are the main inputs to MMLB for debloating the container. The bloat of the container is determined by comparing the original container's files with the debloated container's files.
Once we obtain a correctly functioning container, we define bloat as follows.
\begin{definition}
    Let $F_c$ be the set of files in the original container $c$ and $F_{c'}$ be the set of files in the debloated container $c'$ that are needed for the container to function correctly for a given workload input.
    We have $B_c=F_c-F_{c'}$, where $B_c$ is a set of different files between $F_c$ and $F_{c'}$. We call $B_c$ the bloat of container $c$, namely \textbf{container bloat}.
\end{definition}

Hence, we consider bloat to be the set of files removed from the original container, which are unnecessary for the performed task. We note that this set of files might change if the workload changes. This means that the debloated container is not \emph{general}~\cite{xin2022studying}. 
Thus, general-purpose containers with diverse functionalities and continuous integration require better debloating tools that are capable of generalizing to diverse workload sets.
However, for ML workloads, the standard practice is that a container is deployed to perform one single functionality, i.e., a container deployed for training is never used for inference and vice-versa~\cite{MlOPs}. A trained model is first stored in a model repository and then deployed using a container that is aimed towards inference only. Hence, using only inference or training workloads to generate their respective debloated container is sufficient for ML deployments.

\subsection{Analysis Framework} \label{sec:container_level_measurement}
To understand the sources of bloat, MMLB performs a detailed analysis of the bloat found within the containers focusing on container-level bloat analysis, package-level bloat analysis, vulnerability analysis, and package-dependencies analysis. 
The container-level analysis and the vulnerability analysis are general and can be applied to emerging ML frameworks. The package-level analysis and package-dependency analysis can be adapted to handle different package managers. As of now, our work has mostly focused on package managers related to ML systems which limits the results of the package analysis for applications that do not heavily rely on the three package managers (APT, PIP and Conda~\cite{Conda}) integrated into our framework.

\noindent \textbf{Container-Level Analysis}. We first measure how prevalent bloat is in ML containers using container-level analysis.
We define \textbf{Container bloat degree} to measure how bloated a container is. It is with respect to the size of the container.

\begin{definition}
    Given a container $c$ and its debloated container $c'$, the container bloat degree $d_c$ of $c$ is:
    \begin{equation*}
        \label{eq:container_bloat_degree}
        d_c=\frac{s_c-s_{c'}}{s_{c}}
    \end{equation*}
    where $s_c$ is the size of the original container $c$, $s_{c'}$ is the size of the debloated container $c'$, both measured in the same units.
\end{definition}
The larger $d_c$ is the greater the amount of bloat is in container $c$.
For example, if $d_c=0.9$, it means that bloat accounts for 90\% of the size of the original container $c$. This measure is sensitive to the accuracy of the debloating tool, i.e., if the tool removes necessary files for the operation resulting in a non-functioning container, the calculated $d_c$ has no meaning since some of the files removed are not bloat.
\newline
\newline
\noindent \textbf{Package-Level Analysis.}\label{sec:pkg_level_analysis}
Package-level analysis aims to determine where the bloat originates. The container bloat $B_c$ could include thousands of files. We categorize the files by the packages they belong to. Given a container $c$, we find the list of packages $P$ included in $c$. MMLB then identifies the set of files that belong to each package $p$ in $P$. We then calculate the per-package bloat degree using the following definition.
\begin{definition}
    Let $F_p$ denote the set of files which belong to a package $p$. Given a package $p$, the \textbf{package bloat degree $d_p$} of $p$ is:
    \begin{equation*}
        \label{eq:pkg_bloat_degree}
        d_p=\frac{\texttt{size}(F_p\cap{B_c})}{\texttt{size}(F_p)}
    \end{equation*}
    where \texttt{size($S$)} is a function that calculates the total size of all files in a set $S$.
    \label{def:pkg_bloat_degree}
\end{definition}

Package bloat degree $d_p$ is a measure of how bloated a package is relative to the use of the package in container $c$. 
From another perspective, $d_p$ also represents the usage of package $p$, with lower values indicating higher usage (more precisely, a larger part of the total package is required and is not bloat). For example, if $d_p=1$, i.e., $F_p\subseteq{B_c}$, the package $p$ is entirely unused, and all the files of the package $p$ are bloat. In the opposite extreme case, if $d_p=0$, i.e., $F_p\cap{B_c}$ is $\emptyset$, then all the files of the package are used. The case of $0 < d_p < 1$ represents packages that are partially used, with $d_p$  representing the fraction of files not required.
We develop a package analysis tool to detect the packages installed in a container and find $F_p$ for each package $p$ automatically.

In this paper, we mostly focus our analysis on ML-related packages. A common way to install an ML package is via a package manager.
Different operating systems may use different package managers. For example, Ubuntu uses APT, while CentOS uses YUM~\cite{Yum}. 
MMLB is developed on Ubuntu 18.04. Therefore, we focus on three package managers: APT, PIP, and Conda.
Many ML-related packages are managed by APT, such as \texttt{cuDNN} and \texttt{TensorRT}.
PIP on the other hand is the standard package manager for Python and many ML-related packages are managed by PIP, due to the prevalence of Python in the field of machine learning.
Finally, Conda is a management tool for many languages, including Python, and is used to create machine learning development/deployment environments.
Our tool categorizes packages based on the three types of package managers, allowing us to better understand which package managers contributed the most towards bloat for the containers we examined.

In addition to the analysis of packages categorized by package managers, we also categorize
the detected packages based on functionality. We classify the package into one of the following two categories:
\begin{enumerate}
    \item \textbf{ML packages}. These packages provide functionality to pre-process, build, train, serve, monitor, and accelerate ML models. For example, TensorFlow, PyTorch, and \texttt{cuDNN}.
    \item \textbf{Generic packages}. These packages are not ML packages, such as \texttt{curl} and \texttt{bsdutils}.
\end{enumerate}


\noindent \textbf{Vulnerability Analysis.}
Bloat is known to increase the vulnerabilities in software~\cite{qian2019razor,carve,LMCAS,heo2018effective,brown2019less}. One common measure of the security impact of bloat is to compare the number of Common Vulnerabilities and Exposures (CVEs) found in the removed files to the total number of CVEs discovered~\cite{azad2019less,ahmad2021trimmer,pashakhanloo2022pacjam}.  For assessing CVEs in ML containers, it is crucial that the tools can scan for vulnerabilities in shared libraries and Python packages, which are the two main components of ML systems. Two tools capable of scanning shared libraries are Grype~\cite{Grype} and Trivy~\cite{Trivy}. We integrate both of these tools into MMLB.

These two tools utilize online CVE databases, scanning a container's filesystem for CVEs. It is important to note that this approach can only detect \emph{known} vulnerabilities in containers. 
Both tools depend on generating Software-Bill-Of-Materials (SBOM) files~\cite{zahan2023software}---a nested description of software artifact components and metadata---generated from operating system files that are not typically used for any user functionality. The identified operating system files can be then removed by MMLB. Therefore, as part of the framework, we build a container scanner that scans the filesystem of the debloated containers for the CVEs found in the original (bloated) container. Since debloating only removes files, only CVEs in the bloated containers can exist in the debloated image. Hence, our scanner is guaranteed to find the CVEs in the debloated container.
\newline
\newline
\noindent \textbf{Package Dependency Analysis.} To understand how the installed dependencies of different software affect bloat, MMLB performs a package dependency analysis.
We then define the following.
\begin{definition}
    \label{def:pkg_graph}
    Given an ML container $c$, the \textbf{package attribute graph} of $c$ is $G=(P,E)$, where $P$ is a set of packages, and $E$ is a set of directed edges between packages. 
    For each package $p$ in $P$, MMLB determines three attributes: the package bloat degree ($d_p$), the number of vulnerabilities ($|V|$) and, the depth ($D$). Particularly, the depth of a package is defined as the shortest path from a \textbf{directly-accessed package} to the target package. 
    A \textbf{directly-accessed package} is a package with package bloat degree less than 1, i.e., $d_p<1$. Because such a package has at least one file accessed by the task.
    The depth of a directly-accessed package is set to 1. $E=\{(p_i,p_j)|p_i\neq{p_j}\in{P}\}$, where $p_i$ depends on $p_j$ directly.
\end{definition}

Algorithm \ref{alg:graph_creation} illustrates the process of creating the package attribute graph $G$. The graph $G$ is created from a set of directly-accessed packages $\hat{P}$.
For each package $p_i$ in $\hat{P}$, we use the function \texttt{FindDependencies} to find its dependency packages $P_i$.  The edges are then created between $p_i$ and all the packages in $P_i$. The depth of $p_j$ is set by adding 1 to the depth of $p_i$. The newly identified packages are also added to $\hat{P}$ to find their dependency packages.
For each package, we calculate its package bloat degree and number of known vulnerabilities.
In short, the start nodes of a package attribute graph are those directly-accessed packages. We use Breadth-First-Search to find all the direct and transitive dependencies of these directly-accessed packages.
Figure \ref{fig:attribute_graph} in the Appendix shows an example of the package attribute graph.

\RestyleAlgo{ruled}
\SetKwComment{Comment}{/* }{ */}

\begin{algorithm}[hbt!]
    \SetKwFunction{FindDependencies}{FindDependencies}
    \SetKwFunction{GetBloatDegree}{GetBloatDegree}
    \SetKwFunction{CountKnownVulerabilities}{CountKnownVulerabilities}
    \SetKwFunction{ShortestPath}{ShortestPath}
    \caption{Package attribute graph creation}\label{alg:graph_creation}
    \KwData{Container $c$}
    \KwResult{$G=(P,E)$}
    $P \gets \Phi$\;
    $E \gets \Phi$\;
    $\hat{P} \gets \{p|d_p<1\}$ \;

    \For{$p_i\in\hat{P}$}{
        $P_i \gets$ \FindDependencies{$p_i$}\;
        \For{$p_j \in P_i$}{
            Add $(p_i, p_j)$ to E\;
            \If{$p_j.D$ is not set}{
                $p_j.D \gets p_i.D+1$ \;
            }
            Add $p_j$ to $\hat{P}$\;
        }
        $p_i.d_p \gets$ \GetBloatDegree{$p_i$} \;
        $p_i.|V| \gets$ \CountKnownVulerabilities{$p_i$} \;
        Remove $p_i$ from $\hat{P}$\;
        Add $p_i$ to $P$\;
    }

\end{algorithm}
A package attribute graph includes the information about the packages needed to perform a task in a container. Some packages are needed by the task directly (directly-accessed packages), and some packages are needed because of package dependencies. Based on the definition of a package attribute graph, we define the \textbf{package dependency}, PD($p$), and \textbf{package reach}, PR($p$), in line with~\cite{zimmermann2019small}. PD($p$) measures how many packages that $p$ depends on; PR($p$) measures how many packages depend on $p$.

\begin{definition}
    For every package $p\in P$, the \textbf{package dependency} PD($p$) is the set of all the packages that $p$ depends on directly or transitively.
\end{definition}

\begin{definition}
    For every package $p\in P$, the \textbf{package reach} PR($p$) is the set of all the packages that have a direct or transitive dependency on $p$.
\end{definition}

\section{Measuring Bloat in ML Systems}
Our work aims to quantify bloat in ML systems.
We aim to answer the following six Research Questions (RQs):
\begin{itemize}
    \item \textbf{RQ1}: How prevalent is bloat in ML containers?
    \item \textbf{RQ2}: How does debloating affect the runtime performance of ML deployments in terms of accuracy, and training (tuning or response) time?
    \item \textbf{RQ3}: How does bloat affect ML deployment performance?
    \item \textbf{RQ4}: What are the bloat sources in ML containers?
    \item \textbf{RQ5}: How does bloat increase vulnerabilities in ML deployments?
    \item \textbf{RQ6}: What is the impact of package dependency on bloat and vulnerability?
\end{itemize}

\noindent \textbf{Experimental Setup.} We ran our experiments on an AWS EC2 \texttt{g4dn.xlarge} instance with 4 vCPUs, 16GB of RAM and an Nvidia Tesla T4 GPU. The containers run on Ubuntu 18.04 with Docker version 20.10.9.
We tested MMLB with different GPU setups and observed no differences in our results. The MMLB framework is implemented in approximately 3,300 lines of Python code.

\noindent\textbf{Experiments.} To answer the RQs, we ran MMLB on the 15 containers shown in Table~\ref{tab:containers}, producing a debloated version for each of these containers, extracting the filesystems of the debloated containers and analyzing the bloat and vulnerabilities in the original containers. We then ran experiments to verify that the debloated containers preserved the intended functionality and measured both the deployment and runtime performance metrics. We detail our results next.

\subsection{RQ1: Prevalence of Bloat}
\label{sec:rq1}


\begin{table}[htb]
    \begin{center}
        \caption{Sizes of the original and debloated containers.}
        \label{tab:container_size}
        \begin{tabular}{lrrr}
            \toprule
            Container $c$ & Original Size & Debloated Size & $d_c$         \\
            \midrule
            $c_1$         & 6.34GB        & 2.22GB         & 0.65          \\
            $c_2$         & 8.52GB        & 1.74GB         & \textbf{0.80} \\
            $c_3$         & 15.10GB       & 3.28GB         & 0.78          \\
            $c_4$         & 11.30GB       & 2.74GB         & 0.76          \\
            $c_5$         & 4.54GB        & 1.90GB         & 0.58          \\
            $c_6$         & 8.43GB        & 2.25GB         & 0.73          \\
            $c_7$         & 6.34GB        & 2.21GB         & 0.65          \\
            $c_8$         & 8.52GB        & 1.79GB         & 0.79          \\
            $c_9$         & 12.00GB       & 5.86GB         & 0.51          \\
            $c_{10}$      & 6.34GB        & 3.90GB         & 0.39          \\
            $c_{11}$      & 8.52GB        & 1.79GB         & 0.79          \\
            $c_{12}$      & 4.49GB        & 1.94GB         & 0.57          \\
            $c_{13}$      & 8.43GB        & 3.92GB         & 0.53          \\
            $c_{14}$      & 6.34GB        & 4.04GB         & \textbf{0.36} \\
            $c_{15}$      & 11.50GB       & 4.05GB         & 0.65          \\
            \bottomrule
        \end{tabular}
    \end{center}
\end{table}
We start with the container-level analysis to measure the amount of bloat in the containers listed in Table~\ref{tab:containers}. Table \ref{tab:container_size} presents the sizes of the original containers and the sizes of the debloated containers along with the container bloat degrees $d_c$ according to Definition \ref{eq:container_bloat_degree}.
In our experiments, 13 of the 15 containers have bloat degrees higher than 0.5. The maximum bloat degree is 0.80 for $c_2$ (tuning a BERT model using PyTorch), which indicates that the majority of the size of the container is bloat. The minimum bloat degree is 0.36 for $c_{14}$ (training a MaskRCNN model using TensorFlow), which is still considerable. Only two containers ($c_{10}$ and $c_{14}$) have less than 50\% bloat.

\begin{figure*}
    \centering
    \begin{subfigure}{0.36\textwidth}
        \centering
        \includegraphics[width=\textwidth]{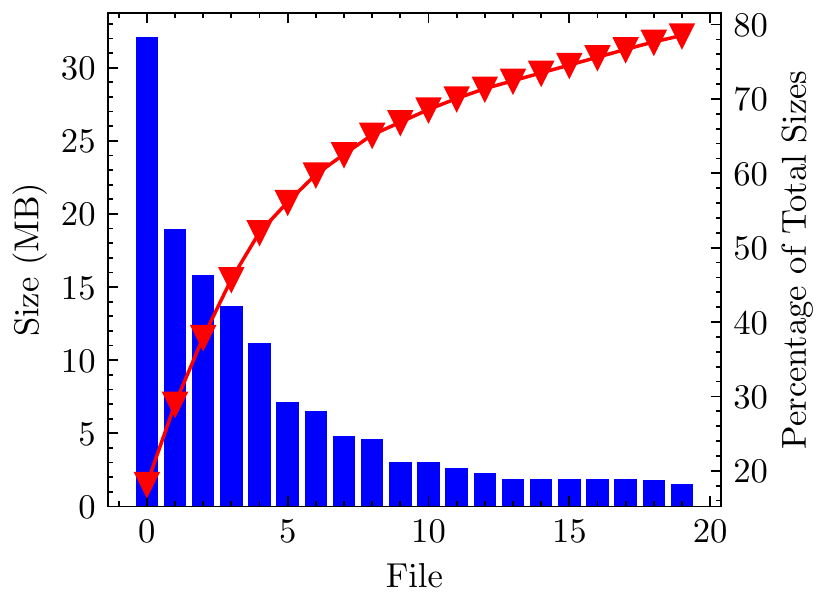}
        \caption{Generic containers}
        \label{fig:generic_common_file_sizes}
    \end{subfigure}
    \hfil
    \begin{subfigure}{0.36\textwidth}
        \centering
        \includegraphics[width=\textwidth]{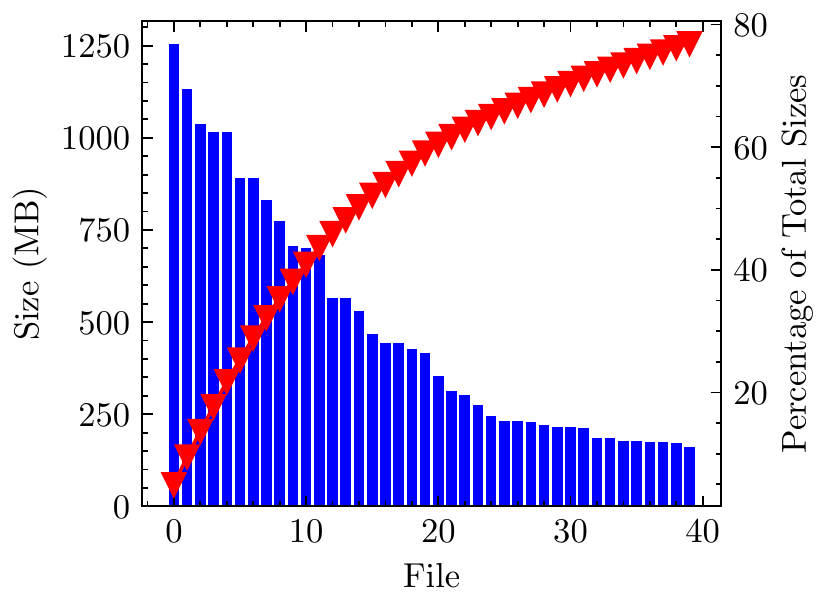}
        \caption{ML containers}
        \label{fig:ml_common_file_sizes}
    \end{subfigure}
    \caption{Pareto charts of files in debloated generic containers and ML containers. The \textit{x-axis} shows indices instead of filenames for simplicity.}
    \label{fig:size_pareto_chart}
\end{figure*}

\begin{table*}[htbp]
    \centering
    \caption{Container bloat degrees of generic containers studied in ~\cite{Rastogi2017Cimplifier}.}
    \label{tab:cimplifier_containers}
    \begin{tabular}{lrrr}
        \toprule
        Container                         & Original Size & Debloated Size & $d_c$ \\
        \midrule
        nginx                             & 133MB         & 6MB            & 0.95  \\
        redis                             & 151MB         & 12MB           & 0.92  \\
        mongo                             & 317MB         & 46MB           & 0.85  \\
        python                            & 119MB         & 30MB           & 0.75  \\
        registry                          & 33MB          & 28MB           & 0.15  \\
        haproxy                           & 137MB         & 10MB           & 0.93  \\
        appcontainers/mediawiki           & 576MB         & 244MB          & 0.58  \\
        eugeneware/docker-wordpress-nginx & 602MB         & 207MB          & 0.66  \\
        sebp/elk                          & 985MB         & 251MB          & 0.75  \\
        \bottomrule
    \end{tabular}
\end{table*}
Table \ref{tab:cimplifier_containers} compares our results with the container bloat degrees of generic containers described in~\cite{Rastogi2017Cimplifier}.
The sizes of the generic containers in Table~\ref{tab:cimplifier_containers} range from 33MB to 985MB, while the sizes of the ML containers in Table~\ref{tab:container_size} range from 4.49GB to 15.10GB.
The sizes of the ML containers are much larger than those of the generic containers.
Meanwhile, the bloat degrees are similar.
Existing file-level debloating tools such as Cimplifier and DockerSlim retain a file as long as it is used by a container, even if only a small fraction of the file is used.
Therefore, if a container accesses a lot of large files that are only partially used, these tools will not be able to debloat the container effectively.
We compared the sizes of files accessed within the generic containers versus the ML containers, sorting their sizes in descending order, and identifying files that collectively contribute to 80\% of the total size of the debloated containers. Figure \ref{fig:size_pareto_chart} shows Pareto charts of these files.
Figure \ref{fig:generic_common_file_sizes} illustrates that up to 20 files in the debloated generic containers account for 80\% of the total size, ranging from 2MB to 33MB. Most of these files are operating systems files that also exist in the ML containers.
On the other hand, Figure \ref{fig:ml_common_file_sizes} shows that up to 40 files of the debloated ML containers contribute to 80\% of the total size, with file sizes ranging from 200MB to 1,200MB, with all files within the debloated ML containers being shared libraries from ML packages (detailed in Table \ref{tab:file_to_pkg} in the Appendix).
Previous research by Agadakos et al.~\cite{agadakos2019nibbler} indicates that shared libraries tend to be bloated.
Given the fact that the shared libraries in ML containers are substantially larger than those found in generic containers, these files exacerbate the overall size of bloat.
Consequently, these file-level debloating tools such as Cimplifier and DockerSlim are less effective in debloating such files in ML containers.
This inadequacy calls for exploring alternative approaches, such as binary-level debloating tools for ML-shared libraries, to address the challenge of efficiently debloating ML containers.

\begin{tcolorbox}
    \textsc{Summary.} Most containers (13 out of 15) contain significant bloat, over 50\% in many cases. Debloated ML containers tend to be larger than generic ones, primarily due to the large sizes of ML-shared libraries. More generic binary-level debloating tools are needed to reduce shared library bloat.
\end{tcolorbox}

%

\subsection{RQ2: Runtime Performance Comparison} \label{sec:rq2}
To understand how removing bloat using Cimplifier affects the runtime of ML containers in terms of loss, accuracy, or training (tuning, serving) time of the containers, we show the performance of three containers running three representative machine learning tasks---that is, training, tuning, and serving--- to validate the performance of the debloated containers versus the original containers.
For tuning and training, we select containers $c_2$ and $c_{10}$.
$c_2$ tunes a pre-trained Bert model on the SQuAD V1.1 dataset~\cite{Rajpurka2016Squachad} for 2 epochs.
$c_{10}$ trains a ResNet model on the ImageNet2012 dataset~\cite{ILSVRC15} for 60 epochs.
For serving, we select $c_{13}$, which serves a ResNet model.
We run the same tasks in the debloated versions of the three containers.
Table \ref{tab:original_debloated_metrics} shows the final loss, accuracy, and training/response time of the original containers and the debloated containers.
The metrics of the original containers and the debloated containers are essentially identical, showing that the runtime of the ML tasks was not affected by the removal of the unused files.

\begin{table}[htb]
    \begin{center}
        \caption{Runtime performance of the original containers and debloated containers. Numbers in the parenthesis represent the metrics of the debloated containers.}
        \label{tab:original_debloated_metrics}
        \begin{tabular}{lcccc}
            \toprule
            Container & Final Loss  & Accuracy    & Training/Response Time    \\
            \midrule
            $c_2$     & 1.72 (1.72) & 84\% (84\%) & 232 minutes (230 minutes) \\
            $c_{10}$  & 2.84 (2.84) & 66\% (66\%) & 13 days (13 days)         \\
            $c_{13}$  & --          & 75\% (75\%) & 1.1 seconds (1.1 seconds) \\
            \bottomrule
        \end{tabular}
    \end{center}
\end{table}

\begin{tcolorbox}
    \textsc{Summary}. Correct debloating does not affect the runtime of the debloated containers compared to the bloated ones.
\end{tcolorbox}
\subsection{RQ3: Deployment Performance Comparison}\label{sec:rq3}
As previously discussed, containers are usually hosted on a container registry.
When a container is deployed, a user needs to pull (i.e., download) the container from a registry and deploy it.
The time taken to pull a container from a registry and run the container is called the \textit{provisioning time}, and consists of three stages:
(1) \textbf{pulling time} denotes the time to download a container from the registry;
(2) \textbf{creation time} denotes the time to create a container that has been already pulled to local storage; this includes the time to create a new layer on top of the container's existing layered filesystem;
(3) \textbf{startup time} denotes the time to start a container that has been already created; this includes the time to mount the created layer to the container's existing layered filesystem.
Note that we do not include the time to execute the container's entrypoint in the startup time for this experiment because it has been measured in \S\ref{sec:rq2}.
We now study the effect of debloating on container provisioning times.
To facilitate a realistic experiment, we upload the original and debloated containers and serve them from the Amazon Elastic Registry.
We use a \texttt{g4dn.xlarge} AWS instance to pull, create and start the containers from the registry.
For each container, we repeat the experiment 10 times.

\begin{figure*}[htbp]
    \centering
    \begin{subfigure}[b]{0.33\textwidth}
        \includegraphics[width=\textwidth]{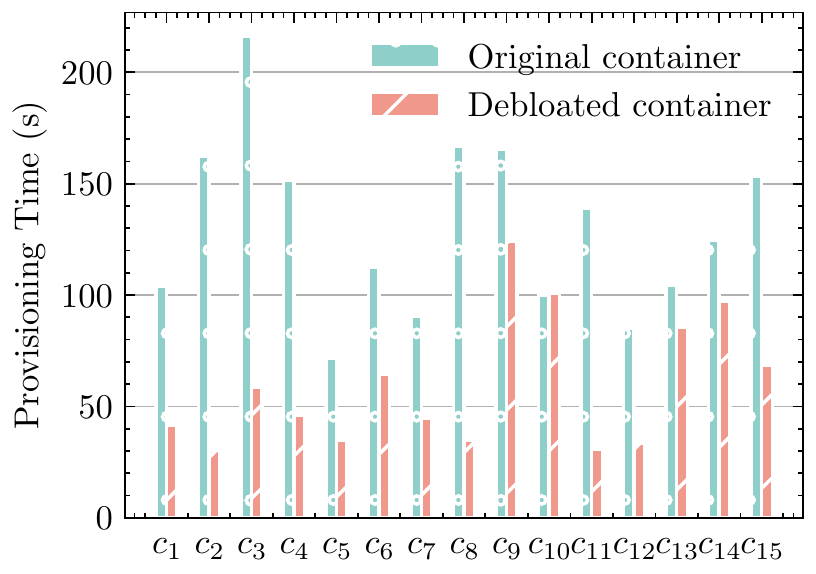}
        \caption{Provisioning time.}
        \label{fig:provisioning_time}
    \end{subfigure}
    \hfil
    \begin{subfigure}[b]{0.33\textwidth}
        \includegraphics[width=\textwidth]{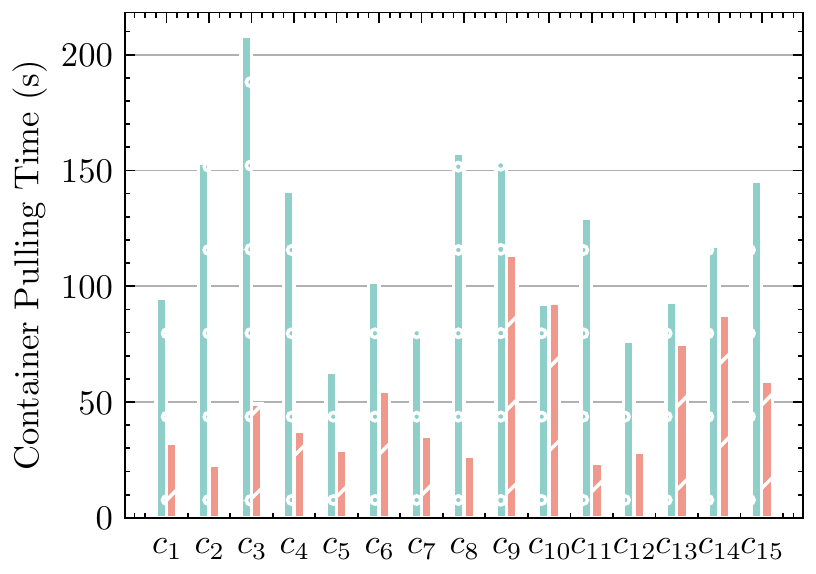}
        \caption{Pulling time.}
        \label{fig:pulling_time}
    \end{subfigure}
    \vskip\baselineskip
    \begin{subfigure}[b]{0.33\textwidth}
        \includegraphics[width=\textwidth]{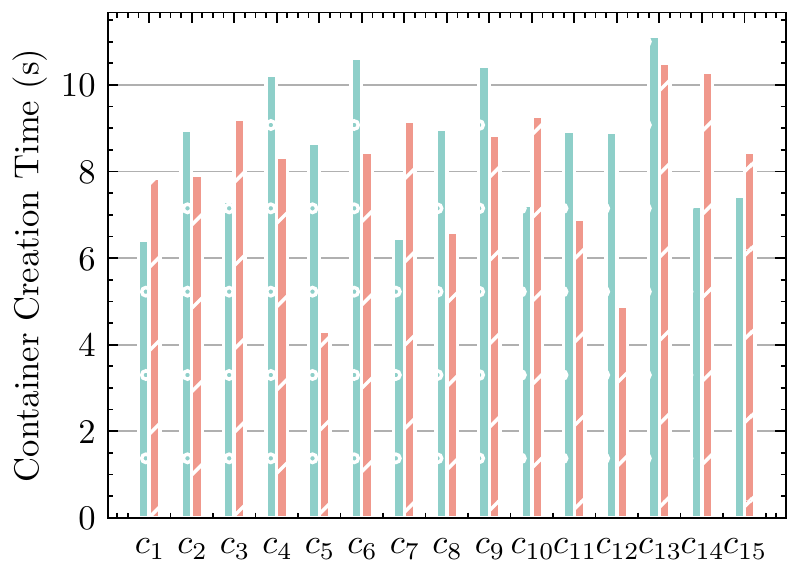}
        \caption{Creation time.}
        \label{fig:creation_time}
    \end{subfigure}
    \hfil
    \begin{subfigure}[b]{0.33\textwidth}
        \includegraphics[width=\textwidth]{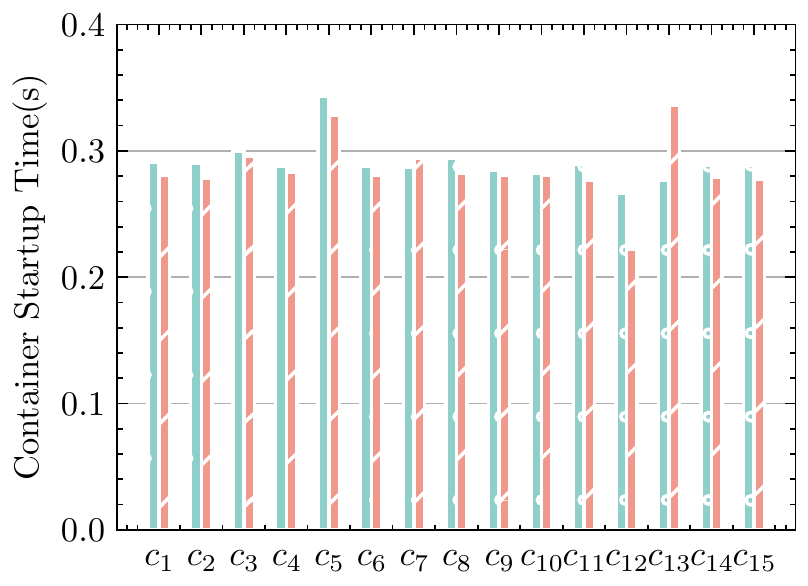}
        \caption{Startup time.}
        \label{fig:startup_time}
    \end{subfigure}
    \caption{Comparing container provisioning time before and after debloating. The provisioning time in (a), along with container pulling time in (b), creation time in (c), and startup time in (d), represent median values calculated over 10 experiments.}
    \label{fig:provisioning_results}
\end{figure*}

Figure \ref{fig:provisioning_time} shows the median provisioning time of the original and debloated containers.
The provisioning time of all the debloated containers is reduced significantly compared to the original ones.
For example, for container $c_3$, the provisioning time reduces from 216s to 59s, a $3.7\times$ acceleration.
To understand which stages of provisioning are affected by bloat,
we also plot pulling time, creation time, and startup time in figures \ref{fig:pulling_time}, \ref{fig:creation_time} and \ref{fig:startup_time}, respectively.
The pulling time of all the debloated containers is much lower than that of the original containers, because of their smaller sizes.
The creation time of the debloated containers is lower or higher than that of the original containers.
As mentioned earlier, creating a new container involves creating a new layer of the existing layered filesystem.
The performance of this process is not directly affected by image sizes.
Other factors, such as dirty pages in memory and network bandwidth can affect the creation time.
Under different scenarios, the creation performance of the debloated container could be better or worse than that of the original one.
The startup time of the debloated containers is essentially identical to that of the original containers.

\begin{tcolorbox}
    \textsc{Summary}. The provisioning time is improved by up to $3.7\times$ after debloating, especially during container pulling.
    In addition to increased latency, transferring bloat wastes resources such as network bandwidth and energy.
    Hosting large ML containers on systems with limited resources, such as tiny machine learning (TinyML) which performs machine learning at the edge, is impractical.
    A better deployment model is needed for these systems.
\end{tcolorbox}


\subsection{RQ4: Sources of Bloat}\label{sec:rq4}
We now shift our focus to the identification of the main sources of bloat for each container using package-level analysis. To understand the packages in the ML containers and container bloat, we divide the packages into different categories according to two different dimensions: package managers and package functionality.

Table \ref{tab:bloat_degree_by_pkg_type} presents the package bloat degrees $d_p$ of the three package sources, PIP, APT and Conda, across all the containers analyzed. APT packages have the highest number of packages at 1,051. PIP packages have the second-highest number of packages at 538. Conda packages have the fewest packages, 203. All of the Conda packages have a bloat degree of 1.
They are all considered bloat in the containers examined. The distributions of the different package types in Table \ref{tab:bloat_degree_by_pkg_type} can be useful in identifying areas for optimization.
\begin{table*}[htbp]
    \begin{center}
        \caption{Bloat degrees categorized by package types. Q(N)=N-th quartile of $d_p$.}
        \label{tab:bloat_degree_by_pkg_type}
        \begin{tabular}{lrrrrr}
            \toprule
            Package Type & Number of Packages & Average $d_p$ & Q1   & Q2   & Q3   \\
            \midrule
            PIP          & 538                & 0.79          & 0.53 & 1.00 & 1.00 \\
            APT          & 1,051              & 0.86          & 0.96 & 0.99 & 1.00 \\
            Conda        & 203                & 1.00          & 1.00 & 1.00 & 1.00 \\
            \bottomrule
        \end{tabular}
    \end{center}
\end{table*}

\begin{figure}[h]
    \centering
    \begin{subfigure}{0.36\textwidth}
        \includegraphics[width=\textwidth]{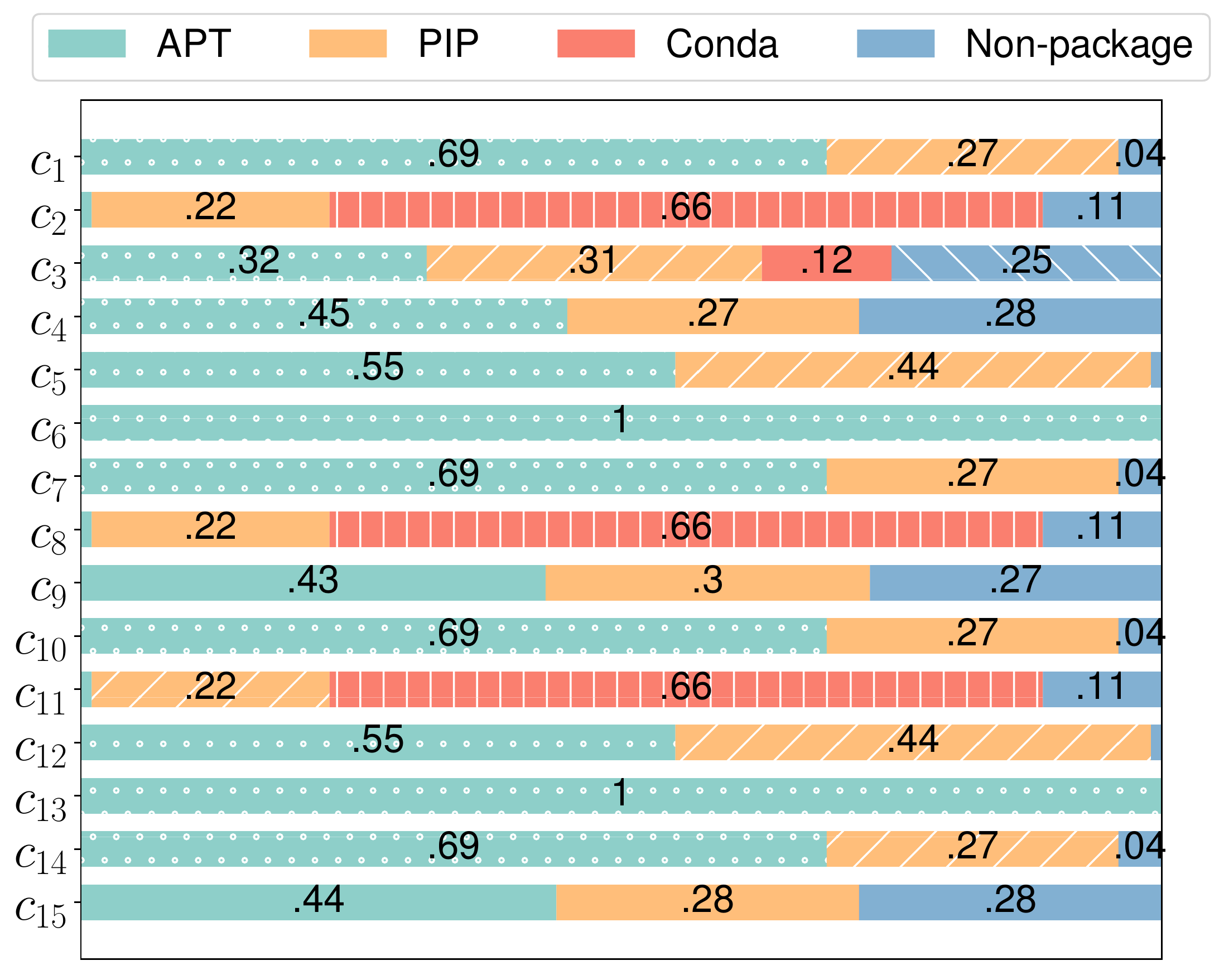}
        \caption{\small{Files in containers.}}
        \label{fig:pkg_dist_original_container}
    \end{subfigure}
    \hfil
    \begin{subfigure}{0.36\textwidth}
        \includegraphics[width=\textwidth]{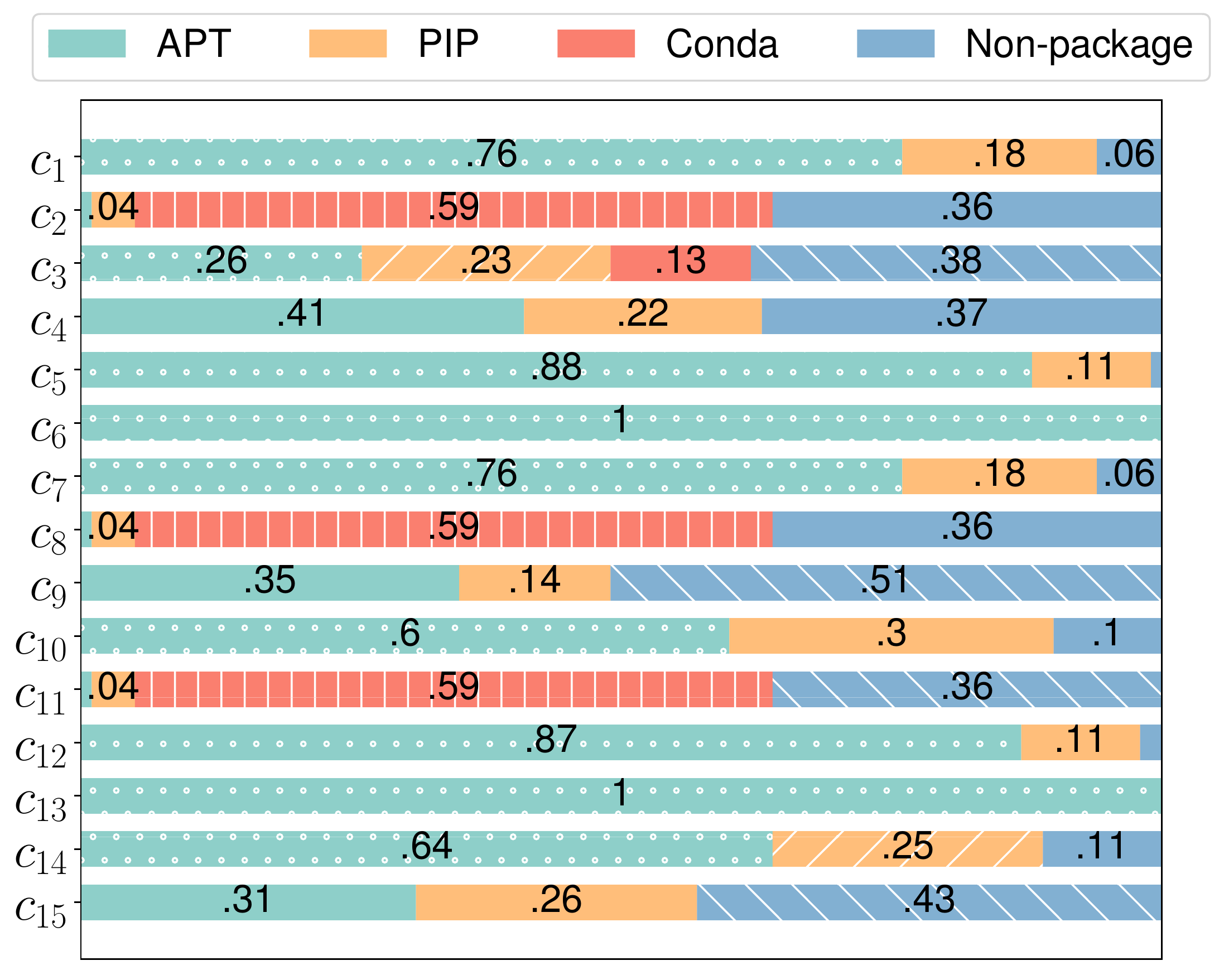}
        \caption{\small{Files in container bloat.}}
        \label{fig:pkg_dist_bloat}
    \end{subfigure}
    \caption{Files in containers and container bloat, categorized according to package managers. Percentages less than 0.03 are omitted.}
    \label{fig:dist_pkg_manager}
\end{figure}

To better understand the sources of container bloat, we categorize files in the containers and container bloat into four categories: APT package files, PIP package files, Conda package files, and Non-package files.
Figure \ref{fig:pkg_dist_original_container} shows the size proportions of the four categories relative to the container sizes.
APT, PIP, and Conda packages account for 72\% to 99\% of the container sizes. This indicates that APT, PIP, and Conda packages are the main components in all the containers we analyzed. Figure \ref{fig:pkg_dist_bloat} shows the size proportions of the four categories relative to the bloat sizes. APT, PIP, and Conda packages account for a significant portion of the bloat size, with more than 50\% of the bloat coming from APT, PIP, or Conda packages in all containers except $c_9$, as shown in Figure \ref{fig:pkg_dist_bloat}.

\begin{figure}[h]
    \centering
    \begin{subfigure}{0.36\textwidth}
        \includegraphics[width=\textwidth]{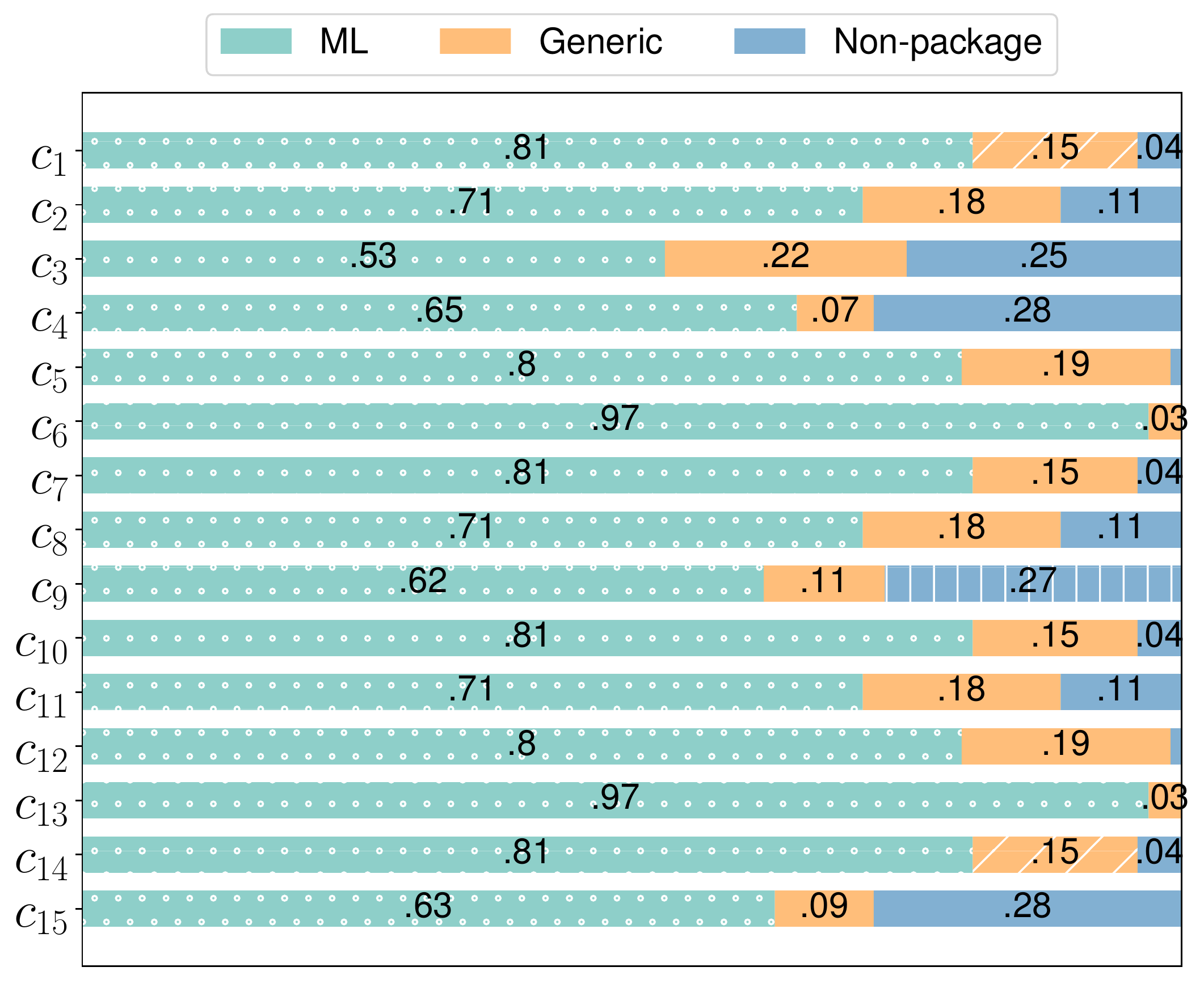}
        \caption{Files in containers.}
        \label{fig:dist_ml_size}
    \end{subfigure}
    \hfil
    \begin{subfigure}{0.36\textwidth}
        \includegraphics[width=\textwidth]{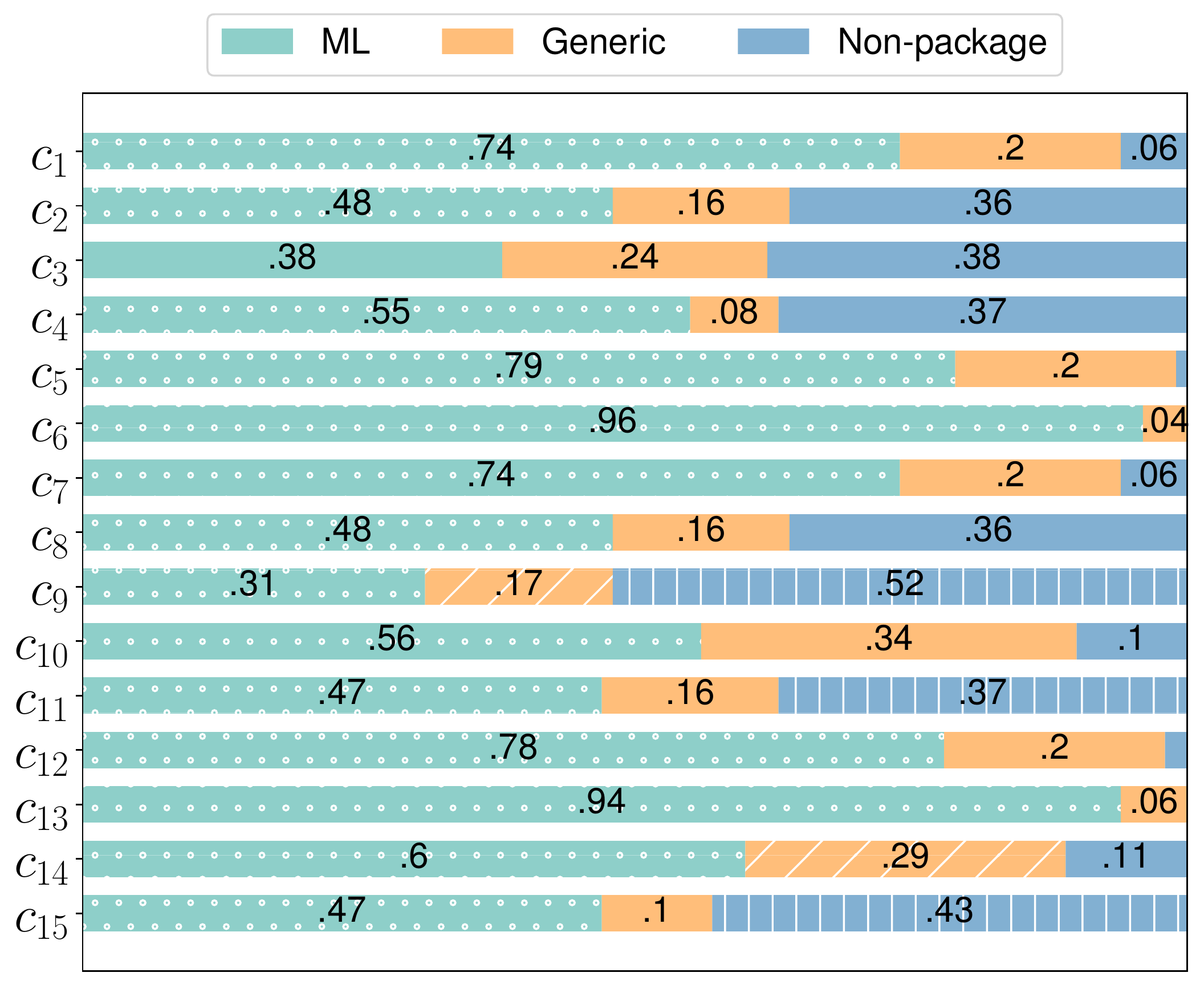}
        \caption{Files in container bloat.}
        \label{fig:dist_ml_bloat_size}
    \end{subfigure}
    \caption{Files in containers and container bloat, categorized according to package functionality. Percentages less than 0.03 are omitted.}
    \label{fig:dist_ml_gpu}
\end{figure}

\begin{figure*}
    \centering
    \begin{subfigure}{0.27\textwidth}
        \includegraphics[width=\textwidth]{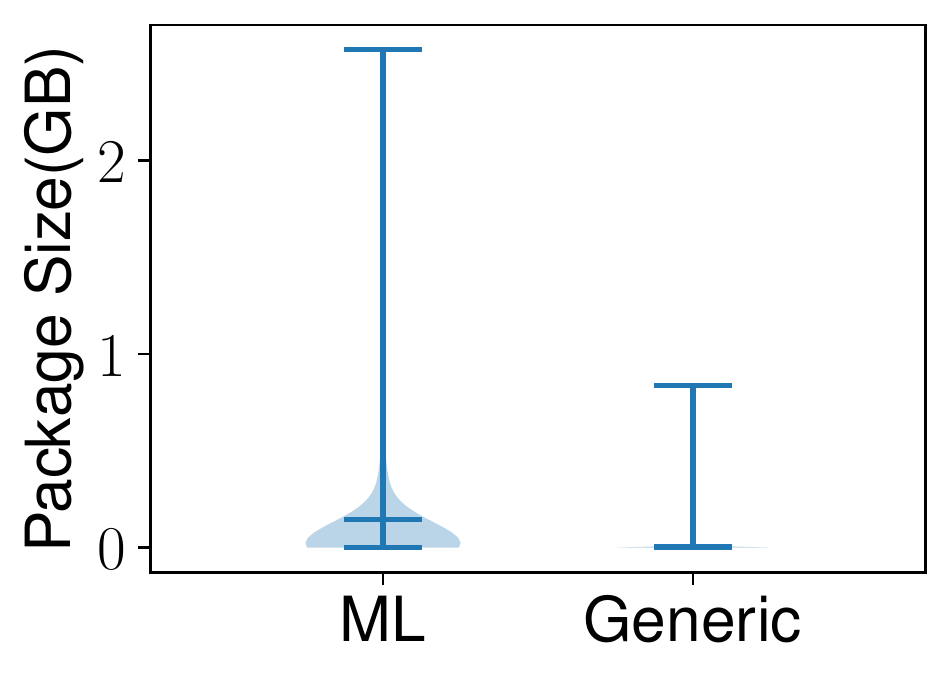}
        \caption{Package sizes.}
        \label{fig:ml_size}
    \end{subfigure}
    \hfill
    \begin{subfigure}{0.28\textwidth}
        \includegraphics[width=\textwidth]{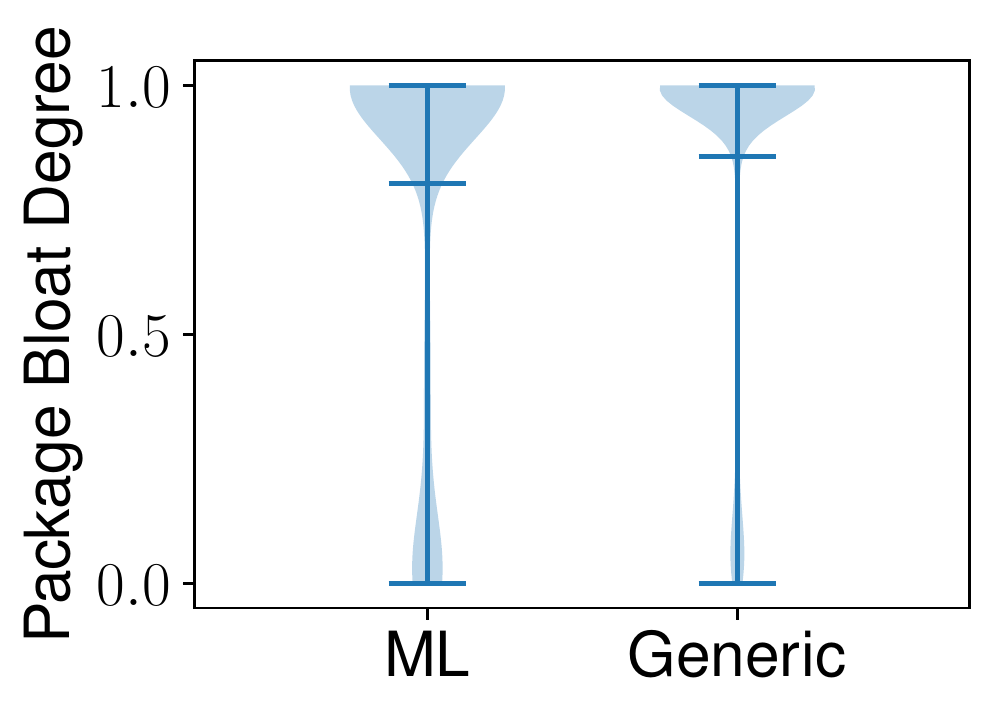}
        \caption{Package bloat degrees.}
        \label{fig:ml_bloat_degree}
    \end{subfigure}
    \hfill
    \begin{subfigure}{0.27\textwidth}
        \includegraphics[width=\textwidth]{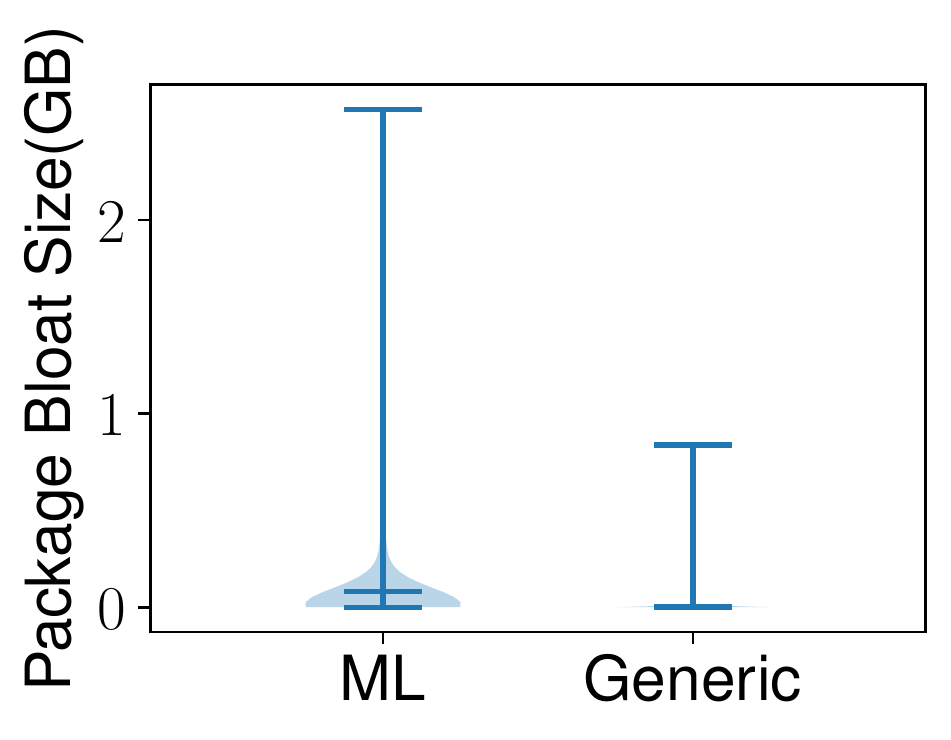}
        \caption{Package bloat sizes}
        \label{fig:ml_bloat_size}
    \end{subfigure}
    \caption{Violin plots of package sizes, bloat degrees, and bloat sizes of ML and Generic packages. The width of the violin at various levels indicates the density of data points there. A wider section of the violin means a higher density of data points at that value.
        The marker in the middle of the violin indicates the median value of the data.
        The two markers at the end of the violin indicate the minimum and maximum values of the data.}
    \label{fig:dist_ml}
\end{figure*}

We also categorize the files in the debloated containers and in the bloat into three categories according to package functionality: ML package files, Generic (i.e., non-ML) package files, and Non-package files.
Figure \ref{fig:dist_ml_gpu} shows the components in containers and container bloat.
As illustrated in Figure \ref{fig:dist_ml_size},
ML packages account for 53\% to 97\% of the container sizes, denoting that they are a major component of ML containers.
Figure \ref{fig:dist_ml_bloat_size} shows the components in container bloat. ML packages account for 31\% to 96\%, signifying that ML packages are a major source of ML container bloat.
Figure \ref{fig:dist_ml} includes three violin plots, showing the distribution of package sizes, package bloat degrees, and bloat sizes of ML and Generic packages.
As can be seen in Figure \ref{fig:ml_size}, ML packages have larger average and maximum sizes than those of Generic packages.
Figure \ref{fig:ml_bloat_degree} displays the distribution of package bloat degrees.
The package bloat degrees of ML packages and Generic packages are similar.
However, ML packages manifest larger average and maximum bloat sizes compared to Generic packages, incurring more bloat than Generic packages, as shown in Figure \ref{fig:ml_bloat_size}.
We find that the top 30 packages sorted by package bloat sizes are all ML packages.
Specifically, 27 of the 30 packages are GPU-related packages.
The details of these packages are listed in Table \ref{tab:top_bloated_pkgs} in the Appendix along with the most frequent unnecessary packages of ML and Generic packages, in Tables \ref{tab:top_ml_pkgs}, and \ref{tab:top_other_pkgs} in the Appendix.

\begin{figure*}[htbp]
    \centering
    \includegraphics[scale=0.27]{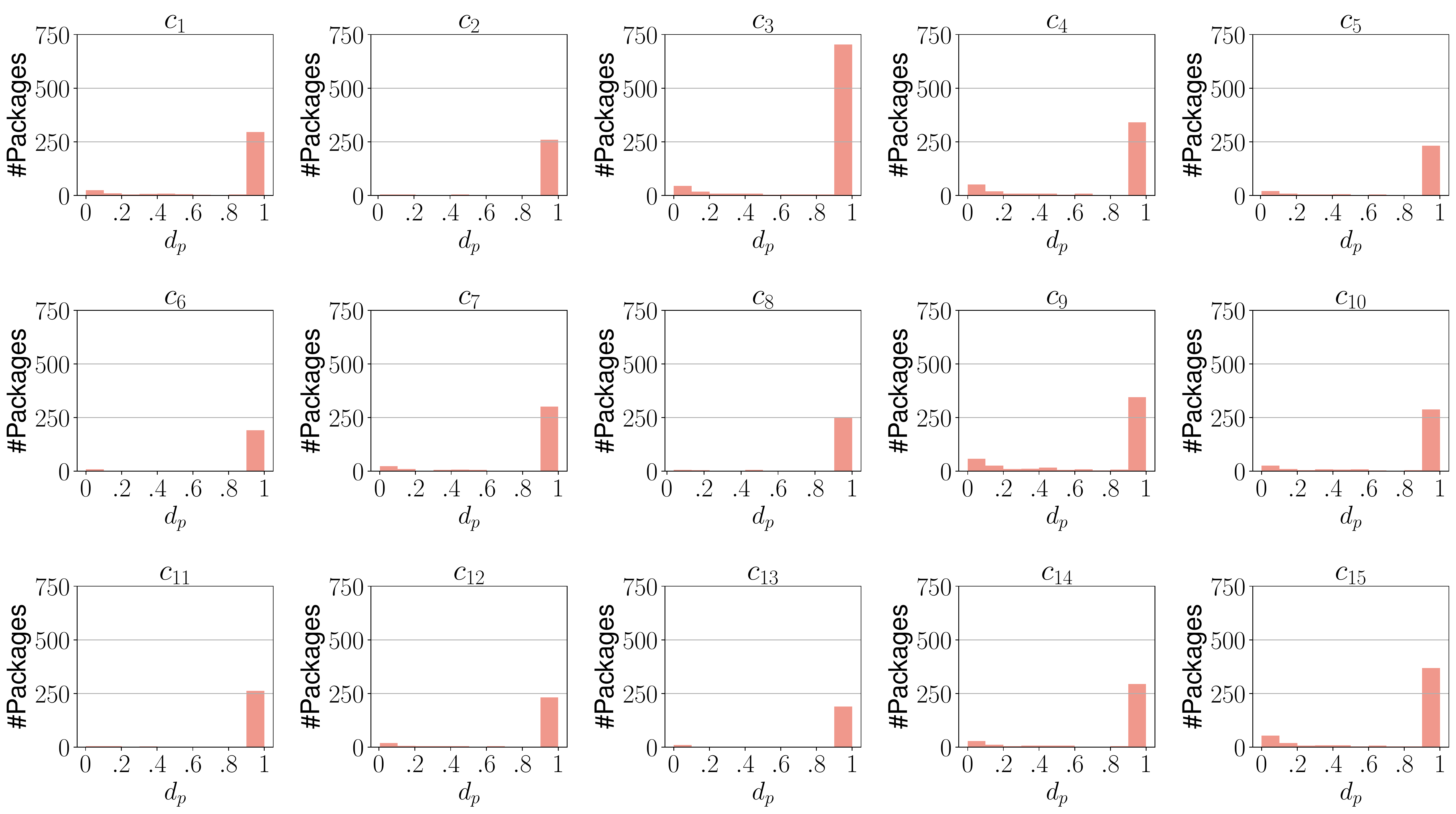}
    \caption{Numbers of packages with different bloat degrees in the containers.
        The \textit{x-axis} is package bloat degree. The bin width of $0.1$.}
    \label{fig:pkg_bloat_degree_dist}
\end{figure*}

\begin{figure*}[htbp]
    \centering
    \includegraphics[scale=0.24]{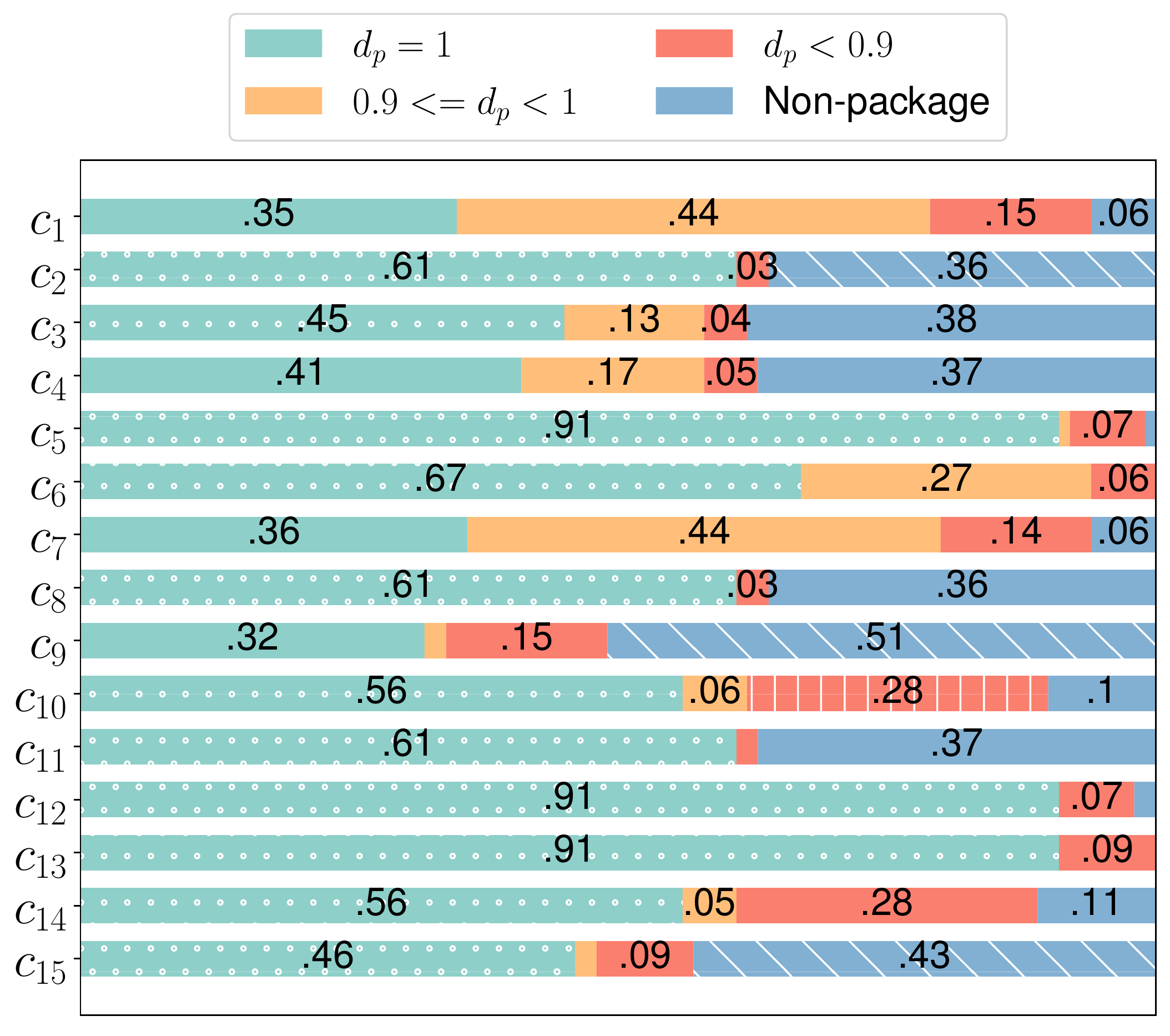}
    \caption{\small{Components in container bloat, divided by package bloat degrees. Percentages less than 0.03 are omitted.}}
    \label{fig:proportion_bloat_degree}
\end{figure*}

We also analyze package bloat degrees for each debloated container and the removed bloat.
Figure \ref{fig:pkg_bloat_degree_dist} shows that all containers include large amounts of packages with bloat degrees ranging from 0.9 to 1. This denotes that the majority of packages installed in ML containers are not used. Another interesting observation is that there is also a reasonable quantity of bloat degrees ranging from 0 to 0.2, such as $c_3$,$c_4$,$c_9$ and $c_{15}$.
This implies that most packages in containers are either used a lot ($0\leq{d_p}<0.2$) or used very little ($0.9\leq{d_p}\leq{1}$).
As for container bloat,
Figure \ref{fig:proportion_bloat_degree} shows that the packages with high bloat degrees ($0.9\leq{d_p}\leq{1}$) account for 50\%+ of the bloat sizes of 13 of the 15 containers.


\noindent\textbf{Case study:}  We now look deeper into the bloat of \textit{libcudnn} as an example package. \textit{libcudnn} is a popular GPU-acceleration library with primitives for deep neural networks~\cite{CuDNN}.
In our experiments, \textit{libcudnn}, which is installed as an APT package, is detected in 12 out of the 15 containers. In containers like $c_1$, $c_3$, $c_4$, $c_6$ and $c_7$, the bloat degree of \textit{libcudnn} is higher than 0.99, which means that less than 1\% of the size of \textit{libcudnn} is used in these containers.
The size of \textit{libcudnn} is around 1.55GB, but only a shared library file of 155KB in size of the package is mostly used. The name of the file is \textit{libcudnn.so.8.*},
which is a shim layer between the application layer and the \texttt{cuDNN} code according to the \textit{libcudnn} documentation\footnote{https://docs.nvidia.com/deeplearning/cudnn/api/index.html}.
In the other containers, a larger fraction of the library is used.
For package users, the issue is a trade-off between functionality and package size. A more modular design or a more dynamic method of packaging is needed to solve this issue.

\begin{tcolorbox}
    \textsc{Summary.} Packages (APT, PIP, and Conda) are the main source of bloat in ML containers. Packages with high bloat degree ($d_p\geq0.9$) contribute to at least 32\% of bloat sizes in all containers. Research is needed to modularize ML packages such that only the fraction of the package needed is included in a container.
\end{tcolorbox}


\subsection{RQ5: Vulnerabilities and Bloat}\label{sec:rq5}
\begin{figure*}
    \centering
    \begin{subfigure}{0.42\textwidth}
        \includegraphics[width=\textwidth]{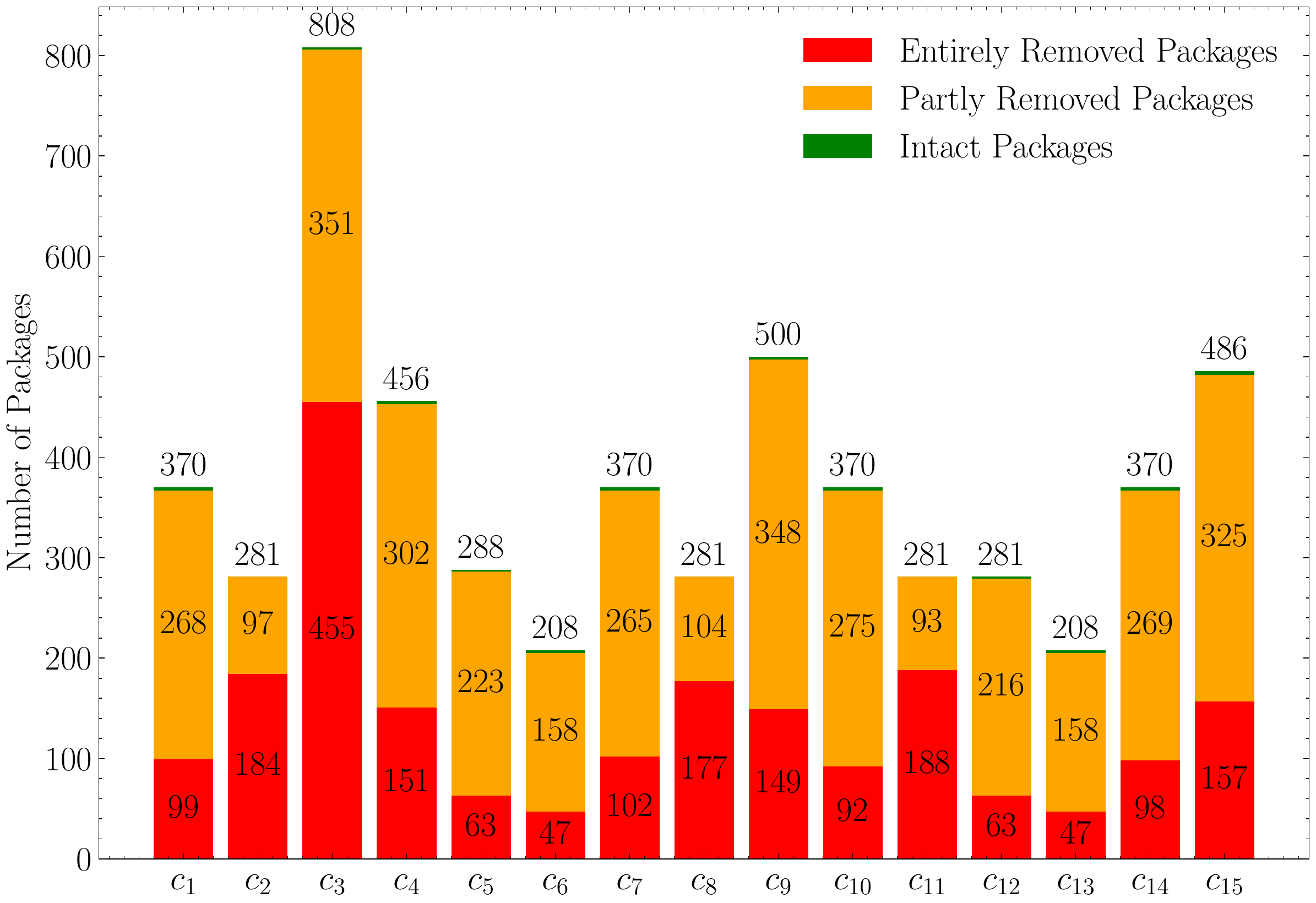}
        \caption{Numbers of packages after debloating the containers.}
        \label{fig:pkg_num_change}
    \end{subfigure}
    \hfill
    \begin{subfigure}{0.42\textwidth}
        \includegraphics[width=\textwidth]{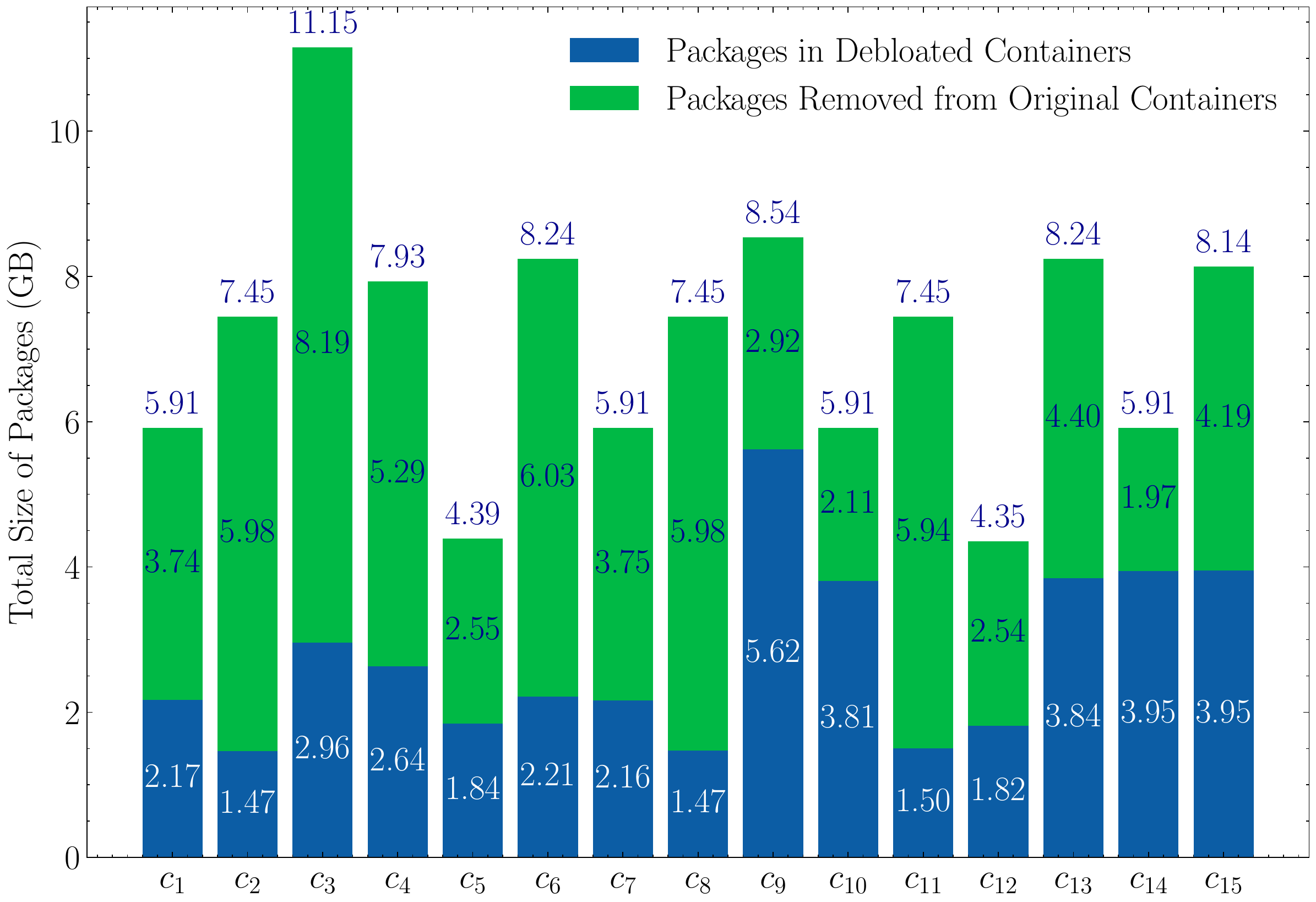}
        \caption{Sizes of packages after debloating the containers.}
        \label{fig:pkg_size_change}
    \end{subfigure}
    \caption{Package reduction after debloating. At the top of each bar in (a) is the total number of packages in the original containers. The numbers of intact packages are omitted because they are too small. At the top of each bar in (b) is the total size of packages in the original containers.}
    \label{fig:analyze_pkg}
\end{figure*}

In this section, we measure how bloat introduces vulnerabilities.
We first analyze changes of packages after debloating.
As shown in Figure \ref{fig:pkg_num_change}, there are a lot of packages in the containers that are entirely removed, i.e., completely unnecessary.
For the remaining packages, only a subset of files are needed. Very few packages remain intact after debloating.
Removal of the unnecessary packages and files reduces the size of the containers noticeably, as shown in Figure \ref{fig:pkg_size_change}.
Table \ref{tab:cves} shows the number of vulnerabilities, and their severity---with the vulnerabilities in the debloated container between parentheses.
Both Trivy and Grype reported similar numbers of CVEs in the original containers.
However, since Grype reported a slightly higher count of CVEs than Trivy, we only show the results of Grype.

For all the containers, the total number of CVEs is reduced significantly after debloating. In the best case ($c_6$ and $c_{13}$), 99\% of CVEs are removed after debloating, with the lowest percentage being for $c_4$ and $c_{15}$ where 66\% of the vulnerabilities are removed.
We see that while bloat increases vulnerabilities, a higher bloat degree does not necessarily translate to a higher number of vulnerabilities found and removed. Take $c_{13}$ as an example, which has a relatively low container bloat degree (0.53), but a high CVE reduction (99\%). This points to the need for a more intelligent container ecosystem where containers are dynamically rebuilt and tested such that the containers with known vulnerabilities are not available to be pulled from a registry.

\begin{figure*}
    \centering
    \begin{subfigure}{0.33\textwidth}
        \includegraphics[width=\textwidth]{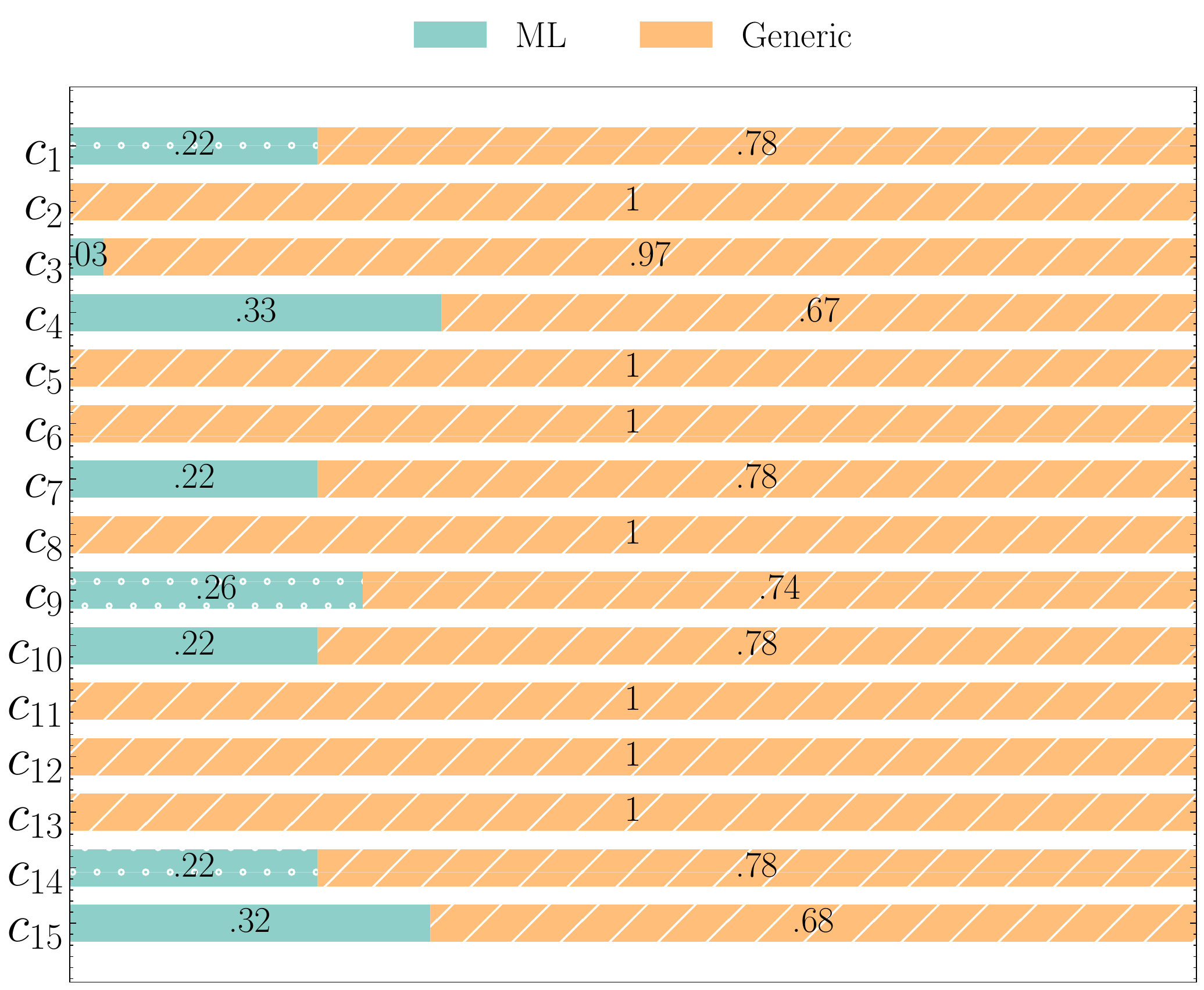}
        \caption{Distribution of reported CVEs.}
        \label{fig:ml_gpu_cve_dist}
    \end{subfigure}
    \hfil
    \begin{subfigure}{0.36\textwidth}
        \includegraphics[width=\textwidth]{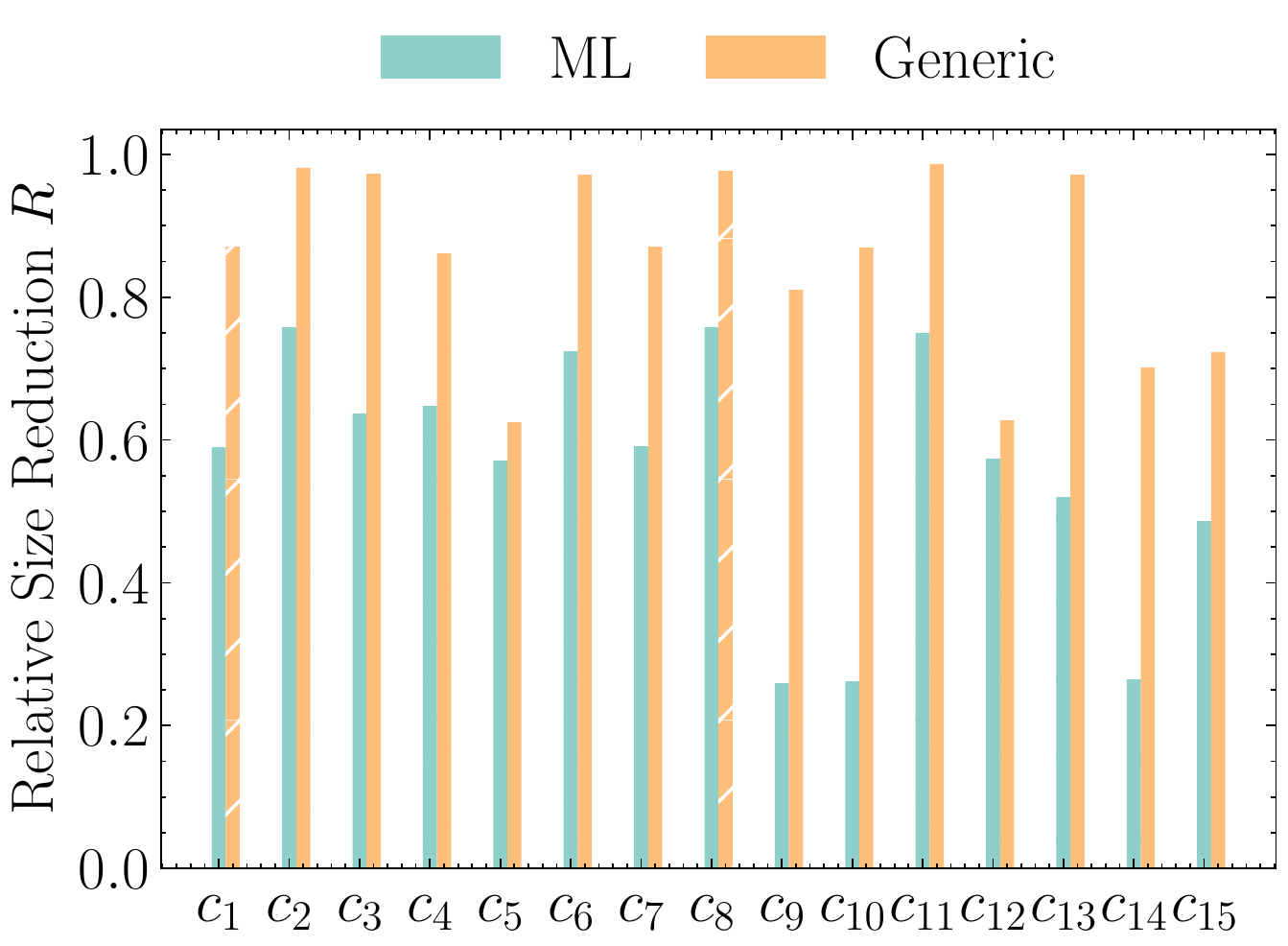}
        \caption{Size reduction of packages.}
        \label{fig:ml_gpu_size_reduction}
    \end{subfigure}
    \caption{Analysis of CVEs in the containers studied.}
    \label{fig:analyze_cve}
\end{figure*}

Hence, we analyze the distribution of CVEs from different package types categorized according to functionality.
Figure \ref{fig:ml_gpu_cve_dist} illustrates the distribution of CVEs from ML and Generic packages in the original containers.
It reveals that Generic packages are the main source of reported CVEs in ML containers.
Combining information from Figure \ref{fig:dist_ml_size} and Figure \ref{fig:ml_gpu_cve_dist}, we found that although ML packages are the main component of ML containers,
very few CVEs are reported for ML packages.
This discrepancy between package sizes and the number of reported CVEs in ML packages is the main reason that we do not see a correlation between the bloat size and the number of CVEs.
Additionally, it also raises the question if the lower number of reported CVEs in ML packages is due to the fact that they are truly more secure, or if it is due to the possibility that ML packages have not been scrutinized as thoroughly as Generic packages for security vulnerabilities.

\begin{small}
    \begin{table*}[htbp]
        \begin{center}
            \caption{Number of CVEs at each severity level found in the original containers and debloated containers.
                Numbers in the parenthesis represent the numbers of CVEs found in the debloated containers. }
            \label{tab:cves}
            \scalebox{0.9}{
                \begin{tabular}{lrrrrrrrrrrrrrrr}
                    \toprule
                    Container & \multicolumn{2}{c}{Critical} & \multicolumn{2}{c}{High} & \multicolumn{2}{c}{Medium} & \multicolumn{2}{c}{Low} & \multicolumn{2}{c}{Negligible} & \multicolumn{2}{c}{Total} & Reduction & $d_c$                                          \\
                    \midrule
                    $c_1$     & 13                           & (12)                     & 159                        & (99)                    & 976                            & (213)                     & 394       & (98)  & 70 & (3) & 1,612 & (425) & 74\% & 0.65 \\
                    $c_2$     & 4                            & (3)                      & 41                         & (10)                    & 207                            & (7)                       & 84        & (0)   & 54 & (0) & 390   & (20)  & 95\% & 0.80 \\
                    $c_3$     & 66                           & (54)                     & 233                        & (135)                   & 1,077                          & (140)                     & 391       & (14)  & 69 & (3) & 1,836 & (346) & 81\% & 0.78 \\
                    $c_4$     & 41                           & (27)                     & 381                        & (231)                   & 1,526                          & (427)                     & 606       & (210) & 57 & (2) & 2,611 & (897) & 66\% & 0.76 \\
                    $c_5$     & 1                            & (1)                      & 27                         & (12)                    & 440                            & (23)                      & 164       & (7)   & 61 & (3) & 693   & (46)  & 93\% & 0.58 \\
                    $c_6$     & 0                            & (0)                      & 51                         & (0)                     & 531                            & (1)                       & 239       & (7)   & 66 & (0) & 887   & (8)   & 99\% & 0.73 \\
                    $c_7$     & 13                           & (12)                     & 159                        & (99)                    & 976                            & (213)                     & 394       & (98)  & 70 & (3) & 1,612 & (425) & 74\% & 0.65 \\
                    $c_8$     & 4                            & (4)                      & 41                         & (27)                    & 207                            & (34)                      & 84        & (6)   & 54 & (1) & 390   & (72)  & 82\% & 0.79 \\
                    $c_9$     & 36                           & (28)                     & 309                        & (205)                   & 1,337                          & (402)                     & 374       & (17)  & 56 & (3) & 2,112 & (655) & 69\% & 0.51 \\
                    $c_{10}$  & 13                           & (12)                     & 159                        & (99)                    & 976                            & (213)                     & 394       & (98)  & 70 & (3) & 1,612 & (425) & 74\% & 0.39 \\
                    $c_{11}$  & 4                            & (3)                      & 41                         & (12)                    & 207                            & (13)                      & 84        & (0)   & 54 & (0) & 390   & (28)  & 93\% & 0.79 \\
                    $c_{12}$  & 1                            & (1)                      & 27                         & (11)                    & 439                            & (14)                      & 164       & (5)   & 61 & (2) & 692   & (33)  & 95\% & 0.57 \\
                    $c_{13}$  & 0                            & (0)                      & 51                         & (0)                     & 531                            & (1)                       & 239       & (7)   & 66 & (0) & 887   & (8)   & 99\% & 0.53 \\
                    $c_{14}$  & 13                           & (12)                     & 159                        & (99)                    & 976                            & (213)                     & 394       & (98)  & 70 & (3) & 1,612 & (425) & 74\% & 0.36 \\
                    $c_{15}$  & 41                           & (28)                     & 381                        & (231)                   & 1,547                          & (429)                     & 615       & (210) & 57 & (2) & 2,641 & (900) & 66\% & 0.65 \\
                    \bottomrule
                \end{tabular}
            }
        \end{center}
    \end{table*}
\end{small}

Figure \ref{fig:ml_gpu_size_reduction} illustrates the relative size reduction $R$ of ML and Generic packages. $R$ is calculated by the equation $R=\sum_{p \in s}(d_p\times\texttt{size}(F_p)) / \sum_{p \in s}\texttt{size}(F_p)$, where $\mathcal{S}\in\{ML, Generic\}$. $ML$ is the set of ML packages, and $Generic$ is the set of Generic packages. It is evident from the figure that for all the containers, Generic packages have a relatively high size reduction, leading to a significant decrease in CVEs. This is because, as seen in Figure \ref{fig:ml_gpu_cve_dist}, most CVEs are from Generic packages. The considerable reduction in the size of Generic packages results in a significant reduction in CVEs.
This explains the low bloat degree but high CVE reduction of $c_{13}$.
As shown in Figure \ref{fig:ml_gpu_size_reduction}, more than 95\% of the Generic packages in $c_{13}$ are removed, while less than 60\% of ML packages are removed.
Table \ref{tab:c13_many_cve} displays the top 10 removed packages in $c_{13}$ sorted by the number of CVEs reported. These packages are all Generic packages with small sizes but a large number of CVEs, with the single package \textit{linux-libc-dev} containing 462 reported CVEs.
On the contrary, very few CVEs are reported for the ML packages as shown in Table \ref{tab:ml_pkg_excerpt}.

\begin{table*}[htbp]
    \begin{center}
        \caption{Top 10 packages sorted by number of CVEs in $c_{13}$. }
        \label{tab:c13_many_cve}
        \begin{small}
            \begin{tabular}{llrrr}
                \toprule
                Name              & Version                  & Size(KB) & Category & \#CVE \\
                \midrule
                linux-libc-dev    & 4.15.0-118.119           & 4,895    & Generic  & 462   \\
                libssl1.1         & 1.1.1-1ubuntu2.1~18.04.6 & 3,457    & Generic  & 18    \\
                libldap-common    & 4.45+dfsg-1ubuntu1.6     & 25       & Generic  & 16    \\
                binutils-common   & 2.30-21ubuntu1~18.04.4   & 6        & Generic  & 10    \\
                perl-modules-5.26 & 5.26.1-6ubuntu0.3        & 18,152   & Generic  & 5     \\
                tar               & 1.29b-2ubuntu0.1         & 480      & Generic  & 4     \\
                libsepol1         & 2.7-1                    & 690      & Generic  & 4     \\
                gnupg-l10n        & 2.2.4-1ubuntu1.3         & 264      & Generic  & 3     \\
                gpgconf           & 2.2.4-1ubuntu1.3         & 338      & Generic  & 3     \\
                gnupg-utils       & 2.2.4-1ubuntu1.3         & 418      & Generic  & 3     \\
                \bottomrule
            \end{tabular}
        \end{small}
    \end{center}
\end{table*}

\begin{table*}[htbp]
    \begin{center}
        \caption{Summary of CVEs in ML packages.}
        \label{tab:ml_pkg_excerpt}
        \begin{small}
            \begin{tabular}{llrrrrr}
                \toprule
                Name         & Version  & Critical & High & Medium & Low & Negligible \\
                \midrule
                tensorflow   & 2.6.0+nv & 2        & 59   & 153    & 4   & 0          \\
                torch        & 1.12.1   & 1        & 0    & 0      & 0   & 0          \\
                scipy        & 1.6.3    & 1        & 0    & 2      & 0   & 0          \\
                horovod      & 0.22.1   & 0        & 1    & 0      & 0   & 0          \\
                onnx         & 1.10.1   & 0        & 1    & 0      & 0   & 0          \\
                transformers & 4.16.2   & 0        & 0    & 1      & 0   & 0          \\
                \bottomrule
            \end{tabular}
        \end{small}
    \end{center}
\end{table*}

\begin{tcolorbox}
    \textsc{Summary.}  Debloating removed up to 99\% of the known CVEs. Reducing the bloat also reduces the attack surface. Generic packages are the major source of reported vulnerabilities. Very few CVEs are reported for ML packages. Greater scrutiny is needed on ML packages to identify their vulnerabilities.
\end{tcolorbox}

\subsection{RQ6: Impact of Package Dependency and Package Reach}\label{sec:rq6}
To measure the impacts of package dependency and package reach on bloat and vulnerabilities, we perform package dependency and reach analysis for PIP and APT packages based on the package attribute graphs we generated\footnote{The attribute graph analysis is not conducted on Conda packages because we found that all the Conda packages have a bloat degree of $1$, which means they are not used.}.
First, we generate a package attribute graph for each container. Then for each package $p$, we calculate the number of packages in PD($p$) ($|PD|$) and the number of packages in PR($p$) ($|PR|$) for each package in the graphs.
As previously discussed,
PD($p$) measures how many packages that $p$ depends on; PR($p$) measures how many packages depend on $p$.
For example, if package $p_0$ depends on package $p_1$ and $p_2$, and is depended on by package $p_3$, $p_4$ and $p_5$, then $|PD|$ of $p_0$ is 2 and $|PR|$ of $p_0$ is 3.

\begin{figure}[h]
    \centering
    \begin{minipage}[t]{0.6\textwidth}
        \centering
        \begin{subfigure}[t]{0.44\textwidth}
            \includegraphics[width=\textwidth]{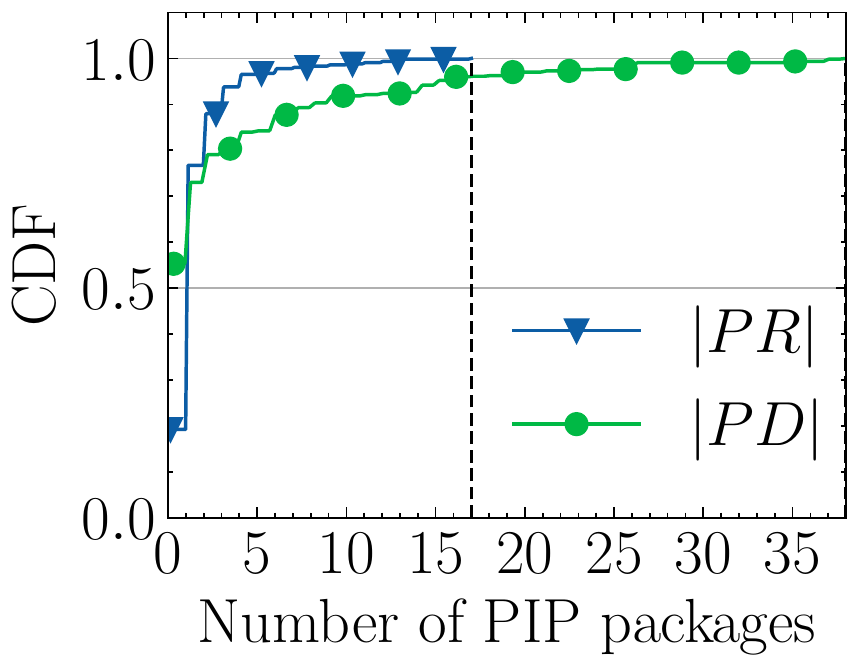}
            \caption{\small{PIP packages.}}
            \label{fig:pip_cdf}
        \end{subfigure}
        \hfill
        \begin{subfigure}[t]{0.44\textwidth}
            \includegraphics[width=\textwidth]{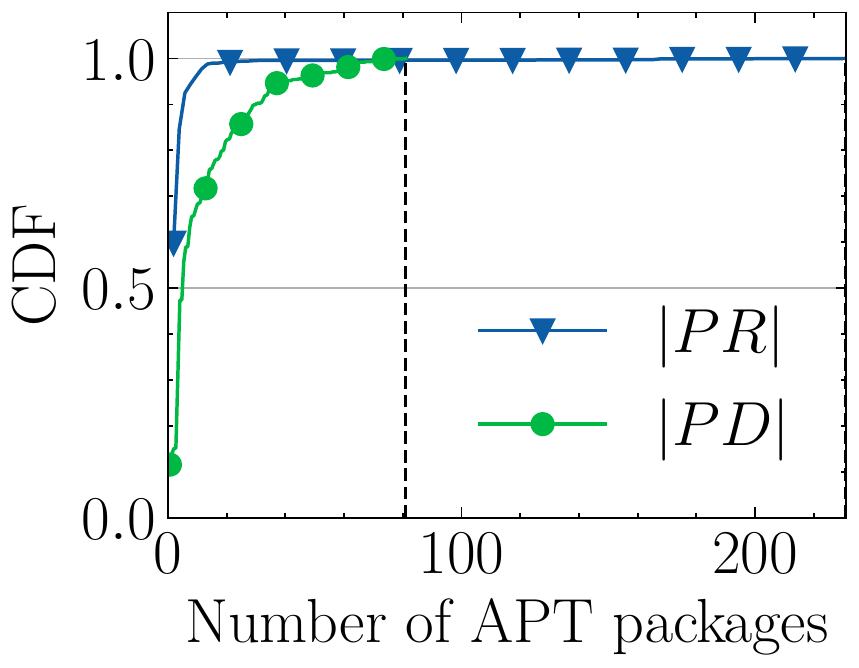}
            \caption{\small{APT packages.}}
            \label{fig:apt_cdf}
        \end{subfigure}
        \caption{Cumulative distribution function of $|PD|$ and $|PR|$.}
    \end{minipage}
    \hfill
    \begin{minipage}[t]{0.32\textwidth}
        \centering
        \includegraphics[width=\textwidth]{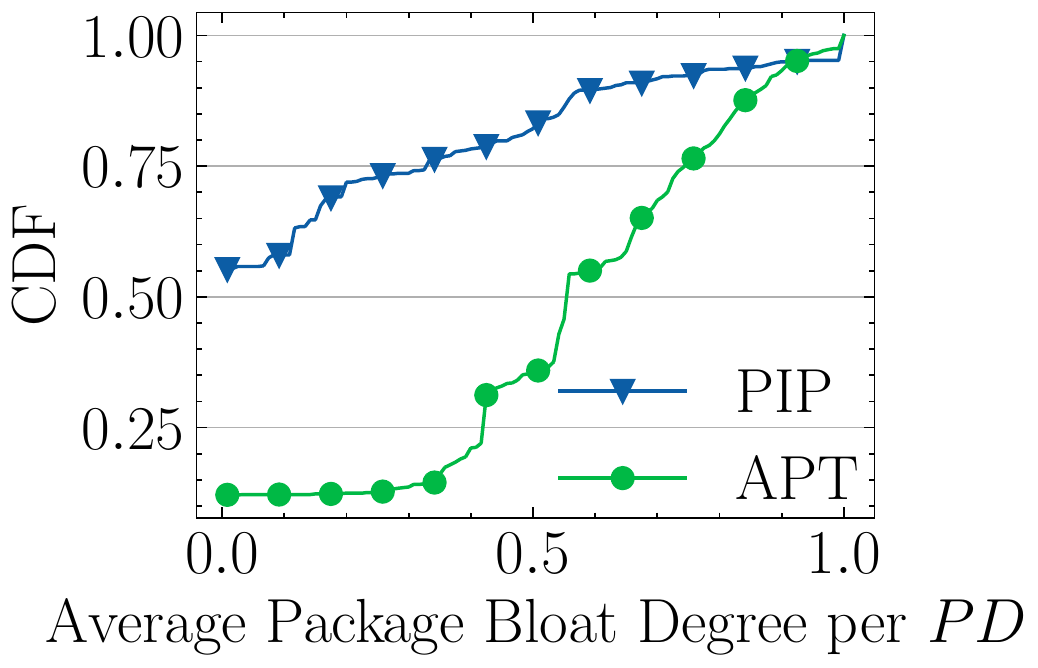}
        \caption{Cumulative distribution function of average package bloat degree.}
        \label{fig:pip_apt_dp}
    \end{minipage}
\end{figure}

Figures \ref{fig:pip_cdf} and \ref{fig:apt_cdf} display the distributions of $|PD|$ and $|PR|$ for PIP and APT packages, respectively. The Cumulative Distribution Function (CDF) reveals that the maximum $|PR|$ value for PIP packages is lower than the maximum $|PD|$ value. However, for APT packages, the maximum $|PD|$ value is much higher than the maximum $|PR|$ value. A high $|PD|$ represents a package that relies on many other packages and hence may introduce bloat and increase the vulnerability in ML systems due to the large number of dependencies. Conversely, a high $|PR|$ indicates that a package is relied upon by many other packages.
Debloating a package with a high $|PR|$ value can lead to significant improvements in removing bloat and vulnerabilities in ML systems because it affects many other packages.
On the other hand, removing a package with a high $|PR|$ value seems riskier than removing one with a low $|PR|$ value.
The $|PD|$ and $|PR|$ values of APT packages are all higher than those of PIP packages, which suggests that the dependency graph of APT packages is denser and more complex.

To understand how a package's depth $D$ affects $|PD|$ and $|PR|$, the correlation between package depth with $|PD|$ and $|PR|$ are calculated, respectively. As can be seen in Table \ref{tab:cor_depth_pd_pr}, for PIP packages, Cor($D$,$|PD|$) is -0.57, which implies a negative correlation between package depth with $|PD|$. It denotes that the directly-accessed packages and the packages near them tend to depend on more packages;
while the packages far from the directly-accessed packages tend to depend on fewer packages. However, there's no such apparent pattern for other correlations.

\begin{table*}[htbp]
    \begin{center}
        \caption{Correlation between packages' depths with $|PD|$ and $|PR|$}
        \label{tab:cor_depth_pd_pr}
        \begin{tabular}{lcc}
            \toprule
            Package Type & Cor($D$,$|PD|$) & Cor($D$,$|PR|$) \\
            \midrule
            PIP          & -0.57           & -0.07           \\
            APT          & -0.04           & -0.01           \\
            \bottomrule
        \end{tabular}
    \end{center}
\end{table*}

\begin{figure}[h]
    \centering
    \begin{subfigure}[b]{0.33\textwidth}
        \includegraphics[width=\textwidth]{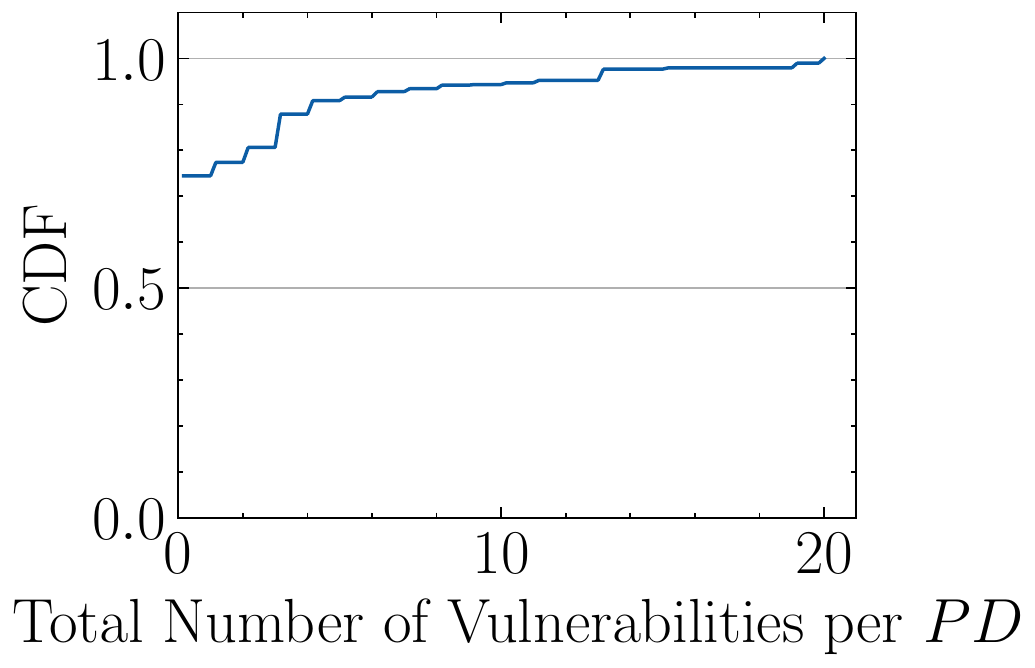}
        \caption{\small{PIP packages.}}
        \label{fig:pip_vul_cdf}
    \end{subfigure}
    \hfil
    \begin{subfigure}[b]{0.33\textwidth}
        \includegraphics[width=\textwidth]{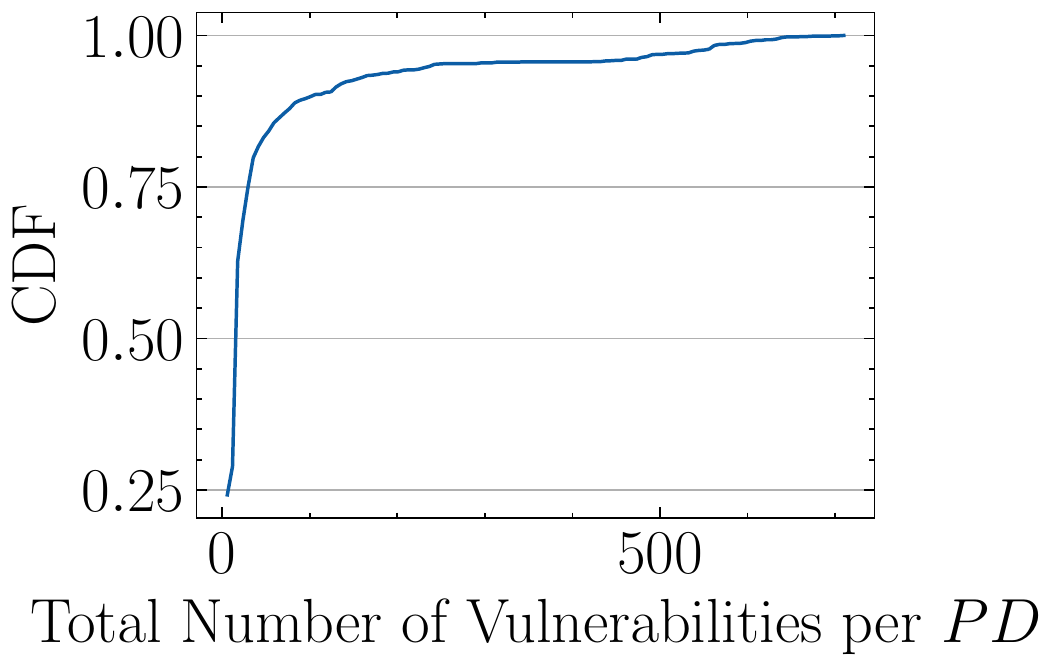}
        \caption{\small{APT packages.}}
        \label{fig:apt_vul_cdf}
    \end{subfigure}
    \caption{Cumulative distribution function of the number of vulnerabilities.}
    \label{fig:subfigures}
\end{figure}

Based on the $PD$ derived from package attribute graphs, we analyze the following attributes: package bloat degree and number of vulnerabilities.
For $PD$ of each package, we calculate the average bloat degree and the total number of vulnerabilities of packages within $PD$.
For example, given a package $p_0$, $PD(p_0)=\{p1,p2\}$, which means $p_0$ depends on $p_1$ and $p_2$.
Assume $p_1$'s package bloat degree is 0.4 and it has 2 vulnerabilities, while $p_2$'s package bloat degree is 0.6 and it has 4 vulnerabilities. Then the average package bloat degree of $PD(p_0)$ is 0.5 ($\frac{0.4+0.6}{2}$); and the total number of vulnerabilities of $PD(p_0)$ is 6 (2+4).
The distributions of average bloat degrees for the PIP and APT packages are displayed in Figure \ref{fig:pip_apt_dp}. The results reveal that, on average, APT packages have a higher dependence on packages with a higher bloat degree compared to PIP packages, as indicated by the majority of the APT package CDF curve lying below the PIP package CDF curve. Figures \ref{fig:pip_vul_cdf} and \ref{fig:apt_vul_cdf} show the CDF of total number of vulnerabilities. APT packages also have more vulnerabilities compared with PIP packages.

\begin{tcolorbox}
    \textsc{Summary.}
    APT packages have higher package dependency and package reach than PIP packages, resulting in a significant source of bloat and vulnerability. In addition, APT packages are reported with more known vulnerabilities. Since APT is used by many other container types, this suggests that more research is needed to understand how our results extend to other containers.
\end{tcolorbox}


\section{Discussion}\label{sec:discuss}

Our results quantify how prevalent bloat is in ML deployments. Bloat not only wastes resources like storage space and network bandwidth to pull the containers and energy used to store and transfer the containers, but also increases the overall security risks to ML deployments. In addition, for container registries and hosting, bloat wastes expensive bandwidth, and increases the provisioning latency, while serving no real purpose in terms of ML runtime performance. We note that while this can be said about many other types of software, ML suffers from increased technical debt due to the way these systems are evolving compared to other applications.
As noted by Sculley et al.~\cite{sculley2015hidden}, ML deployments have only a small fraction of their code performing the actual task. While companies spend numerous resources in time and effort optimizing ML algorithms, we believe that bloat is mostly being overlooked. Increasingly relying on pre-packaged containers, sharing them across organizations and use cases, and oversimplifying the entire deployment process, leads to increased vulnerabilities and resource wastage. Our work is a plea for leaner and more modular ML systems and deployments similar to the call by Wirth over 25 years ago~\cite{wirth1995plea}.

ML packages are a major source of bloat in ML containers. 
However, some of the ML packages, such as GPU-related packages, are not open-source.
Debloating tools that do not require source code as input are needed for these ML packages.
Furthermore, ML packages include many large shared library files, meaning that even if only a small part of the libraries is used, the entire file needs to be retained. This makes existing container debloating tools less effective in debloating ML containers.
To address the issue of bloat in ML containers, a more intelligent packaging and container ecosystem that prevents pulling bloated containers including known vulnerabilities is needed.

\noindent \textbf{Threats to Validity.} Empirical studies inherently face validity threats. We have tried to mitigate these via our experimental design. While we have only studied four Docker base images, these were tested using various workloads (training, tuning, serving), and two ML frameworks (PyTorch and TensorFlow). The chosen ML models also span diverse fields such as NLP, Image Classification, and Image Segmentation. Ultimately, we tested 15 containers.
In addition, to reduce the error due to the variability when measuring the provisioning time, we repeated all experiments ten times and compared the median values.
We used Grype and Trivy for vulnerability analysis, which are two widely used container scanning tools. However, these tools depend on file management that could be eliminated during the debloating process. As a result, we implemented components in our framework to detect vulnerabilities in the debloated containers by locating the specific files linked to each CVE in the vulnerability report and verifying if these files are deleted, suggesting the removal of the CVE.

\section{Related Work}\label{sec:related_work}

Bloat can be studied on three levels, which are source code bloat, binary bloat, and container bloat.

\textbf{Source code bloat} occurs during the development process and is often caused by the inclusion of unused features or outdated configurations. To address this issue, several tools have been proposed to remove bloat from source code directly, i.e., not when the code is deployed. These tools are usually tailored to specific programming languages such as C/C++~\cite{brown2019carve,heo2018effective}, JavaScript~\cite{ye2021jslim} and PHP~\cite{azad2019less}.  These tools typically result in source-code size reduction ~\cite{brown2019carve,heo2018effective,azad2019less}, a reduction in CVEs ~\cite{brown2019carve,heo2018effective,ye2021jslim,azad2019less} and reduction in binary sizes~\cite{brown2019carve}.

\textbf{Binary bloat} studies the presence of unused code and dependencies in a binary, i.e., after the source code is compiled. Several binary debloating tools have been developed. These tools work on software binaries, for example, Linux shared libraries or executables~\cite{tang2021xdebloat,quach2019bloat,qian2019razor,ahmad2021trimmer,chen2018toss,agadakos2019nibbler}, Java jar packages~\cite{dewan2021bloatlibd,ponta2021used,bruce2020jshrink,soto2022coverage}, and Android APK packages~\cite{jiang2018reddroid,tang2021xdebloat}. These tools typically measure their efficacy by studying the source of bloated dependencies~\cite{soto2021comprehensive,soto2021longitudinal}, code size reduction~\cite{quach2019bloat,bruce2020jshrink,tang2021xdebloat,soto2022coverage,agadakos2019nibbler}, binary size reduction~\cite{ahmad2021trimmer,jiang2018reddroid,ponta2021used,chen2018toss,tang2021xdebloat,soto2022coverage}, and CVE reduction ~\cite{quach2019bloat,qian2019razor,ahmad2021trimmer,chen2018toss}. Moreover, Tang et al. ~\cite{tang2021xdebloat} have shown that removing binary bloat can lead to improvements in runtime performance, as well as lower memory and power usage.

\textbf{Container bloat} is the focus of our work.
Container bloat is a phenomenon where unnecessary code, files, and packages are included in containers, mostly due to the way that containers are built today. To address this issue, several container debloating tools have been developed to remove unused files and generate a slimmer container~\cite{Rastogi2017Cimplifier,DockerSlim,thalheim2018cntr}. These tools have been applied to web servers, databases, and web applications. However, their applicability to ML containers has never been studied before, nor has the bloat characteristics of ML containers been investigated.
To bridge these gaps, our work performs container bloat analysis on ML containers. We design and implement a framework to conduct an extensive analysis of container bloat in ML containers, including quantifying container bloat, identifying the sources of the bloat, and assessing the effects on container security and performance.

\section{Conclusion}
We conducted a quantified measurement study of bloat in machine learning containers using MMLB, a framework we developed.
The framework uses container-level analysis, package-level analysis, vulnerability analysis, and package dependency analysis to quantify the amount, the source, the performance overhead, and the vulnerabilities of bloat in ML containers. 
We measured 15 ML containers and found that bloat affects a large number of ML containers, with up to 80\% of some containers consisting of bloat. APT packages are a significant source of bloat and vulnerabilities due to package dependencies. 
Additionally, ML packages are a main source of container bloat.
Bloat increases provisioning time by up to 370\% and increases vulnerabilities by up to 99\%.
However, very few CVEs are reported for these ML packages. Greater scrutiny of ML code for vulnerabilities is needed.
We hope that this work will inspire further research on the quantification of technical debt in machine learning systems, the development of more modular ML systems and a more intelligent container ecosystem.


\bibliographystyle{ACM-Reference-Format}
\bibliography{sample-base}
\clearpage
\appendix
\section{Appendix}\label{sec:appendix_paper}

\subsection{Package Attribute Graph}
\begin{figure}[h]
    \centering
    \includegraphics[scale=0.27]{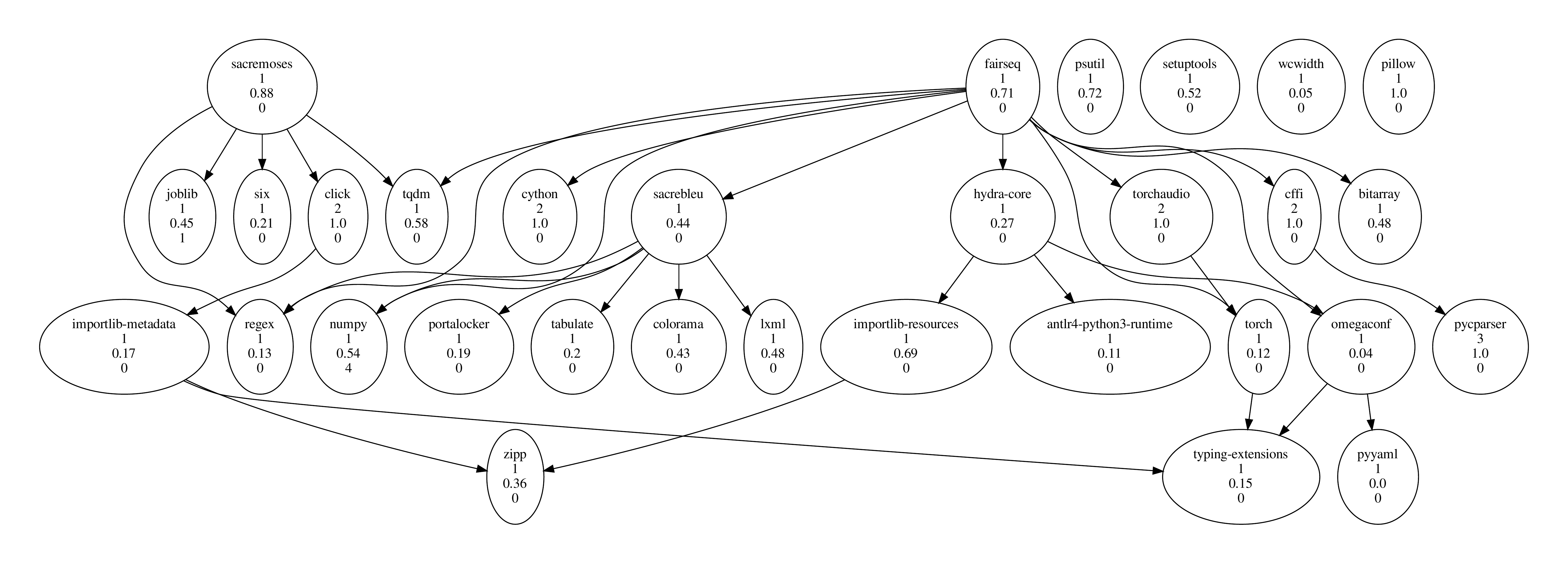}
    \caption{An example package attribute graph. Each node in the graph is a package. The four lines of text in a node are the package name, depth, package bloat degree and number of known vulnerabilities of this package.}
    \label{fig:attribute_graph}
\end{figure}

\subsection{Summary of Large Files in Debloated ML Containers}
Tabled \ref{tab:file_to_pkg} is discussed in \S\ref*{sec:rq1}. It lists the large files found in debloated ML containers. All the files are from ML packages.
\begin{small}
    \begin{table}[h]
        \caption{Summary of the large files in debloated ML containers.}
        \label{tab:file_to_pkg}
        \begin{tabular}{lcccr}
            \toprule
            File Name                         & Package Name     & Package Manager & Package Functionality & Size(MB) \\
            \midrule
            libtorch\_cuda.so                 & Torch            & PIP             & ML                    & 1,253    \\
            libcudnn\_cnn\_infer.so           & libcudnn         & APT             & ML                    & 1,133    \\
            \_pywrap\_tensorflow\_internal.so & TensorFlow       & PIP             & ML                    & 1,038    \\
            libdali\_kernels.so               & nvidia-dali-cuda & APT             & ML                    & 700      \\
            libcusolver.so                    & libcusolver      & APT             & ML                    & 565      \\
            libtorch\_cpu.so                  & Torch            & PIP             & ML                    & 469      \\
            libdali\_operators.so             & nvidia-dali-cuda & APT             & ML                    & 428      \\
            libcudnn\_ops\_infer.so           & libcudnn         & APT             & ML                    & 417      \\
            libcufft.so                       & libcufft         & APT             & ML                    & 353      \\
            libcublasLt.so                    & libcublas        & APT             & ML                    & 314      \\
            \bottomrule
        \end{tabular}
    \end{table}
\end{small}

\subsection{Analysis of Package Bloat}

The following tables are discussed in \S\ref*{sec:rq3}: Tables~\ref{tab:top_bloated_pkgs},~\ref{tab:top_ml_pkgs} and ~\ref{tab:top_other_pkgs}.

\begin{footnotesize}
    \begin{table}[H]
        \begin{center}
            \caption{Top 30 packages sorted by bloat size.}
            \label{tab:top_bloated_pkgs}
            \begin{small}
                \begin{tabular}{lcrrc}
                    \toprule
                    Package Name            & Package Manager & Avg. Bloat Size(MB) & Avg. Bloat Degree & Functionality \\
                    \midrule
                    pytorch                 & Conda           & 2,570.50            & 1.00              & ML                    \\
                    cudatoolkit             & Conda           & 1,568.82            & 1.00              & ML                    \\
                    nvidia-dali-cuda110     & PIP             & 1,012.79            & 0.75              & ML                    \\
                    mkl                     & Conda           & 820.39              & 1.00              & ML                    \\
                    libcudnn8               & APT             & 761.92              & 0.61              & ML                    \\
                    libcudnn8-dev           & APT             & 656.03              & 0.99              & ML                    \\
                    libnvinfer8             & APT             & 388.89              & 1.00              & ML                    \\
                    libnvinfer7             & APT             & 360.26              & 0.75              & ML                    \\
                    libcublas-dev-11-0      & APT             & 341.44              & 1.00              & ML                    \\
                    libcufft-dev-11-0       & APT             & 326.84              & 1.00              & ML                    \\
                    libcusolver-11-0        & APT             & 320.08              & 0.38              & ML                    \\
                    cuda-cusolver-10-2      & APT             & 305.97              & 1.00              & ML                    \\
                    magma-cuda110           & Conda           & 237.68              & 1.00              & ML                    \\
                    libcusolver-11-4        & APT             & 233.57              & 0.52              & ML                    \\
                    nsight-compute-2021.3.0 & APT             & 214.81              & 1.00              & ML                    \\
                    cupy-cuda114            & PIP             & 202.19              & 1.00              & ML                    \\
                    libnpp-11-5             & APT             & 193.96              & 1.00              & ML                    \\
                    libcusolver-11-2        & APT             & 187.30              & 0.25              & ML                    \\
                    libnpp-11-4             & APT             & 184.32              & 1.00              & ML                    \\
                    libcusolver-dev-11-0    & APT             & 172.51              & 1.00              & ML                    \\
                    torch                   & PIP             & 170.74              & 0.10              & ML                    \\
                    cuda-nvgraph-10-2       & APT             & 166.31              & 1.00              & ML                    \\
                    libcusparse-dev-11-0    & APT             & 163.62              & 1.00              & ML                    \\
                    libnpp-11-2             & APT             & 160.68              & 1.00              & ML                    \\
                    libnpp-dev-11-0         & APT             & 158.76              & 1.00              & ML                    \\
                    nsight-compute-2021.2.2 & APT             & 158.24              & 1.00              & ML                    \\
                    cuda-cufft-10-2         & APT             & 150.16              & 1.00              & ML                    \\
                    cuda-npp-10-2           & APT             & 142.92              & 1.00              & ML                    \\
                    nsight-compute-2020.3.0 & APT             & 141.49              & 1.00              & ML                    \\
                    libnpp-11-0             & APT             & 140.40              & 1.00              & ML                    \\
                    \bottomrule
                \end{tabular}
            \end{small}
            \vskip -0.1in
        \end{center}
    \end{table}
\end{footnotesize}

\begin{table}[H]
    \begin{center}
        \caption{Top 5 unnecessary ML packages. }
        \label{tab:top_ml_pkgs}
        \begin{small}
            \begin{tabular}{lcrr}
                \toprule
                Package Name           & Package Manager & Frequency & Avg. Size(MB) \\
                \midrule
                tensorboard-plugin-wit & PIP             & 8         & 3.28          \\
                torchtext              & PIP             & 6         & 19.53         \\
                cuda-nvdisasm-11-0     & APT             & 6         & 27.57         \\
                cuda-gdb-11-0          & APT             & 6         & 14.92         \\
                cuda-sanitizer-11-0    & APT             & 6         & 29.35         \\
                cuda-memcheck-11-0     & APT             & 6         & 0.41          \\
                cuda-cuobjdump-11-0    & APT             & 6         & 0.27          \\
                tensorflow-metadata    & PIP             & 5         & 0.58          \\
                scikit-learn           & PIP             & 5         & 69.84         \\
                sentencepiece          & PIP             & 5         & 2.80          \\
                torchvision            & PIP             & 5         & 47.90         \\
                tf-slim                & PIP             & 5         & 3.37          \\
                tensorflow-datasets    & PIP             & 5         & 14.96         \\
                tensorrt               & PIP             & 4         & 2.53          \\
                libcudnn8-dev          & APT             & 4         & 656.03        \\
                nvidia-dali-cuda110    & PIP             & 3         & 1,303.69      \\
                cudatoolkit            & Conda           & 3         & 1,568.82      \\
                \bottomrule
            \end{tabular}
        \end{small}
        \vskip -0.1in
    \end{center}
\end{table}

\begin{table}[H]
    \begin{center}
        \caption{Top 1 unnecessary Generic packages. .}
        \label{tab:top_other_pkgs}
        \begin{small}
            \begin{tabular}{lccc}
                \toprule
                Package Name       & Package Manager & Frequency & Avg. Size(MB) \\
                \midrule
                gpgv               & APT             & 15        & 0.45          \\
                sensible-utils     & APT             & 15        & 0.02          \\
                libaudit-common    & APT             & 15        & 0.00          \\
                libdb5.3           & APT             & 15        & 1.72          \\
                bsdutils           & APT             & 15        & 0.19          \\
                libsemanage1       & APT             & 15        & 0.26          \\
                libsemanage-common & APT             & 15        & 0.01          \\
                base-passwd        & APT             & 15        & 0.04          \\
                linux-libc-dev     & APT             & 15        & 5.05          \\
                \bottomrule
            \end{tabular}
        \end{small}
        \vskip -0.1in
    \end{center}
\end{table}

Table~\ref{tab:top_bloated_pkgs} lists the top 30 packages sorted by bloat size over the 15 containers, with sizes ranging from 140.40MB to 2,570.50MB.
A package may appear in multiple containers.
The average bloat size and the average bloat degree of each package are shown in the Avg. Bloat Size(MB) and Avg. Bloat Degree columns, respectively.
Many of the packages have average bloat degrees of 1, which means these packages are unnecessary.
All these packages are ML packages.

Table~\ref{tab:top_ml_pkgs} lists the top 5 unnecessary ML packages. Unnecessary packages are packages with a package bloat degree of 1.
The Frequency column shows the number of occurrences of each package as an unnecessary package in the 15 containers.
The packages with the highest 5 frequencies are listed in the table.

Table~\ref{tab:top_other_pkgs} lists the top 1 unnecessary Generic packages.
The Frequency column shows the number of occurrences of each package as an unnecessary package in the 15 containers.
The packages with the highest frequency are listed in the table.
Note that not all packages are displayed.
\end{document}